\documentclass[aps,pre,twocolumn,superscriptaddress,showkeys,amsmath,amssymb,showpacs]{revtex4-1}

\usepackage[latin1]{inputenc}
\usepackage{graphicx}
\usepackage{amsmath,amssymb}
\usepackage{dcolumn}
\usepackage{bm}
\usepackage{epsfig}
\usepackage{longtable}
\usepackage{subfigure}
\usepackage{multirow}
\usepackage{enumerate}
\usepackage{url}
\usepackage{afterpage}
\usepackage{color}

\newcommand{\bfr}{ {\bf r}} 
\newcommand{\bfrp}{ {\bf r'}} 
\newcommand{\bfrpp}{ {\bf r^{\prime\prime}}} 
\newcommand{\bfR}{ {\bf R}} 
\newcommand{\bfG}{ {\bf G}} 
\newcommand{\bfGp}{ {\bf G'}} 
\newcommand{\bfzero}{ {\bf 0}} 
\newcommand{\bfq}{ {\bf q}} 
\newcommand{\bfk}{ {\bf k}} 
\newcommand{\omegap}{\omega^\prime} 

\newcommand{\GnWn}{\ensuremath{G_\text{0}W_\text{0}}\,}

\allowdisplaybreaks

\begin{document}

\title[]{All-electron periodic {\GnWn} implementation with numerical atomic orbital basis functions: algorithm and benchmarks}
\date{\today}

\author{Xinguo Ren\footnote{Corresponding author: renxg@iphy.ac.cn}}
     
      \address{Beijing National Laboratory for Condensed Matter Physics, Institute of Physics, Chinese Academy of Sciences, 3rd South Str. 8, Beijing 100190, China}
\author{Florian Merz}
       \address{Lenovo HPC Innovation Center, Meitnerstr. 9, D-70563 Stuttgart, Germany}
\author{Hong Jiang}
       \address{Beijing National Laboratory for Molecular Sciences, College of Chemistry and Molecular Engineering, Peking University, 100871 Beijing, China}
\author{Yi Yao}
        \address{Department of Chemistry, University of North Carolina, Chapel Hill, North Carolina 27599, USA}
        \address{Department of Mechanical Engineering and Materials Science, Duke University, Durham, North Carolina 27708, USA}
\author{Markus Rampp}
       \address{Max Planck Computing and Data Facility, Giessenbachstrasse 2, D-85748 Garching, Germany}
\author{Hermann Lederer}
       \address{Max Planck Computing and Data Facility, Giessenbachstrasse 2, D-85748 Garching, Germany}
\author{Volker Blum}
       \address{Department of Chemistry, Duke University, Durham, North Carolina 27708, USA}
       \address{Department of Mechanical Engineering and Materials Science, Duke University, Durham, North Carolina 27708, USA}
\author{Matthias Scheffler}
       \address{Fritz-Haber-Institut der Max-Planck-Gesellschaft, Faradayweg~4-6, D-14195 Berlin, Germany}
       \address{Physics Department, Humboldt-Universit\"{a}t zu Berlin, Zum Gro{\ss}en Windkanal 6, D-12489 Berlin, Germany}

\begin{abstract}
We present an all-electron, periodic {\GnWn} implementation within the numerical atomic orbital (NAO) basis framework. A localized variant of
the resolution-of-the-identity (RI) approximation is employed to significantly reduce the computational cost of evaluating and 
storing the two-electron Coulomb repulsion integrals. We demonstrate that the error arising from localized RI approximation can be reduced to
an insignificant level by enhancing
the set of auxiliary basis functions, used to expand the products of two single-particle NAOs. An efficient algorithm is introduced to deal with     
the Coulomb singularity in the Brillouin zone sampling that is suitable for the NAO framework. We perform systematic convergence tests and identify a set of 
computational parameters, which can serve as the default choice for most practical purposes. Benchmark calculations are carried out for a set of prototypical semiconductors
and insulators, and compared to independent reference values obtained from an independent \GnWn implementation based on linearized augmented plane waves (LAPW)
plus high-energy localized orbitals (HLOs) basis set, as well as experimental results. With a moderate (FHI-aims \textit{tier} 2) NAO basis set, our \GnWn calculations
produce band gaps that typically lie in between the standard LAPW and the
LAPW+HLO results. Complementing \textit{tier} 2 with highly localized 
Slater-type orbitals (STOs), we find that the obtained band gaps show an overall convergence 
towards the LAPW+HLO results. The algorithms and techniques developed in this work 
pave the way for efficient implementations of correlated methods within the NAO framework. 
\end{abstract}



\maketitle

\section{\label{sec:introduction}Introduction}
 The electronic band structure determines the behavior of electrons and consequently a large variety of properties of periodic 
 materials. 
 Therefore, accurate computations of electronic band structures are crucial for \textit{ab initio} quantum mechanical descriptions of materials.
 Kohn-Sham (KS) density-functional theory (DFT) \cite{Hohenberg/Kohn:1964,Kohn/Sham:1965} within its local-density \cite{Vosko/Wilk/Nusair:1980,Perdew/Zunger:1981} and generalized gradient approximations (LDA/GGAs) \cite{Becke:1988,Perdew/Burke/Ernzerhof:1996} offers a relatively inexpensive approach to compute band structures 
  in solids, and has significantly improved our understanding of solid materials. However, LDA and GGAs do not include the correct underlying physics to describe experimentally relevant band structures, i.e., electron addition or removal energies and other quasiparticle properties. The substantial underestimation of the band gaps of
  insulating materials, and occasionally the incorrect description of  energy level orderings have been a major drawback of this approach.
 Hybrid density functionals  \cite{Becke:1993}, especially those incorporating a portion of screened exact exchange \cite{Heyd/Scuseria/Ernzerhof:2003},
 have gained increased popularity in condensed matter and materials science community, due to their overall improved band structure description. 
 However, a certain level of empiricism is often required to tune the mixing and screening parameters to suitable values, and such functionals still lack the desired 
 predictive power. Recently there has been progress to determine these semi-empirical parameters automatically in a self-adaptive way 
\cite{Shimazaki/Asai:2008,Marques/etal:2011,Skone/etal:2016,He/Franchini:2017,Cui/etal:2018}.

 An alternative, and formally more rigorous approach to calculate electronic band structures is the Green-function based many-body perturbation theory 
 \cite{Abrikosov/Gorkov/Dzyaloshinski:1975,Fetter/Walecka:1971}, whereby the band structure can be determined from the poles
 of the interacting Green function $G$ of the system. Via the Dyson equation, $G$ is linked to a reference, non-interacting Green function $G_0$ 
  in terms of  a non-hermitian,
 dynamic self-energy, which encompasses all the many-body exchange-correlation effects. In practice, approximations have to be made to calculate 
 the self-energy, and a popular choice has been the so-called $GW$ approximation \cite{Hedin:1965,Strinati/Mattausch/Hanke:1980,Hybertsen/Louie:1986,Godby/etal:1988,
 Aryasetiawan/Gunnarsson:1998,Aulbur/Jonsson/Wilkins:2000,Onida/Reining/Rubio:2002,Rinke/etal:2005,Marom/etal:2012,Golze/Dvorak/Rinke:2019}, whereby the self-energy is given by a product of the
 Green function $G$ and the screened Coulomb interaction $W$. Despite its simplicity, the $GW$ approximation has been shown to yield significantly improved band
 structures, compared to their DFT counterparts, for a wide range of materials. Because the $GW$ band structure is derived from a dynamical self-energy, and
 describes correctly the quasiparticle nature of single-particle excitations in materials, the $GW$ band structure is often called \textit{quasiparticle}
 band structure in the literature.

Since the 1980's,  the $GW$ approach has been implemented within different numerical frameworks, ranging from the pseudopotential
plane wave method \cite{Hybertsen/Louie:1986,Bruneval/Gonze:2008,Deslippe/etal:2012,Govoni/Galli:2015}, the projector augmented
wave (PAW) method 
\cite{Arnaud/Alouani:2000,Lebegue/etal:2003,Shishkin/Kresse:2006,Klimes/Kaltak/Kresse:2014,Hueser/Olsen/Thygesen:2013}, 
a mixed representation of plane waves and real-space grid 
\cite{Rojas/Godby/Needs:1995,GW_space-time_method:1998}, the linearized
muffin-tin orbital (LMTO) method \cite{Aryasetiawan/Gunnarsson:1995}, to the linearized augmented plane wave (LAPW) method 
\cite{Ku/Eguiluz:2002,Friedrich/etal:2009,Friedrich/etal:2010,Jiang/etal:2013,Nabok/Gulans/Draxl:2016} and Gaussian orbital basis sets 
\cite{Rohlfing/Kruger/Pollmann:1993,Wilhelm/Hutter:2017,Wilhelm:2018}.   
Needless to say, each implementation has its own advantages as well as limitations, but the considerable efforts behind these various
implementations are clear indications of the importance and wide-spread influence of the $GW$ method. As a matter of fact, the $GW$ functionality is 
becoming a standard component of many widely used first-principles computational software packages.

In addition to the significant efforts devoted to its numerical implementations, the $GW$ methodology itself is also under active development. In practice, $GW$ calculations
are often done on top of a preceding KS-DFT LDA/GGA calculation, or a hybrid functional calculation within the generalized KS (gKS) scheme.  In this case,
the $GW$ band structure is obtained as a one-shot correction to the KS-DFT one, and such a computational scheme is termed as \GnWn 
in the literature. 
The \GnWn scheme frequently performs very well, but the obtained results obviously depend on the starting point, i.e., 
the approximation used in the preceding (generalized) KS-DFT calculation.
Various schemes beyond the simple \GnWn approach have been proposed and tested, including, e.g., Green-function-only partially 
self-consistent $GW$ (denoted as $GW_0$) \cite{vonBarth/Holm:1996,Shishkin/Marsman/Kresse:2007}, quasiparticle self-consistent $GW$ (QPscGW) \cite{Faleev/Schilfgaarde/Kotani:2004,Schilfgaarde/Kotani/Faleev:2006}, 
and fully self-consistent $GW$  \cite{Ku/Eguiluz:2002,Stan/Dahlen/Leeuwen:2006,Rostgaard/Jacobsen/Thygesen:2010,Caruso/etal:2012,Caruso/etal:2013,Kutepov:2016,Kutepov:2017}, 
but {\GnWn} remains to be the most widely used approach in practical calculations.

The $GW$ approach has been the state-of-the-art electronic-structure method for determining quasiparticle energies in semiconductors and insulators in
the last three decades.  In recent years, $GW$ has become also popular as an approach to determine ionization energies and
electron affinities of molecules \cite{Rostgaard/Jacobsen/Thygesen:2010,Ke:2011,Blase/Attaccalite/Olevano:2011,Foerster/etal:2011,Blase_review:2012,Ren/etal:2012,
Caruso/etal:2012,Bruneval/Marques:2012,Setten/etal:2013,Pham/etal:2013,Knight/etal:2016,Setten/etal:2015,Maggio/etal:2017}. Aside from its considerable importance 
in realistic description of organic materials, applying $GW$ to molecules allows one to systematically benchmark the accuracy across different, independent $GW$ 
implementations, as was done in the $GW$100 test set project \cite{Setten/etal:2015}, and to benchmark the accuracy of the $GW$ method against the traditional 
quantum chemistry approaches such as the coupled cluster (CC) method \cite{Bruneval/Marques:2012,Knight/etal:2016}. Most recently, an in-depth diagrammatic analysis has been carried out to compare 
and contrast $GW$ with the equation-of-motion CC method \cite{Lange/Berkelbach:2018}.

Our own $GW$ implementation \cite{Ren/etal:2012} has been carried out within the all-electron numerical atomic orbital (NAO) based FHI-aims code package 
\cite{Blum/etal:2009,Havu/etal:2009,Ihrig/etal:2015,Levchenko/etal:2015,Knuth/etal:2015}. 
Comprehensive benchmark tests \cite{Lejaeghere/etal:2016,Jensen/etal:2017,Huhn/Blum:2017} indicate that FHI-aims offers impressive precision for ground-state DFT calculations. 
 The $GW$ implementation in FHI-aims was initially done for the \GnWn scheme 
and finite systems for non-periodic boundary condition \cite{Ren/etal:2012}, but soon extended to the fully self-consistent $GW$ 
\cite{Caruso/etal:2012,Caruso/etal:2013} scheme. Like other correlated methods available in FHI-aims, our $GW$ implementation is based on a technique known as variational density fitting \cite{Whitten:1973,Dunlap/Connolly/Sabin:1979} or the
resolution-of-the-identity (RI) approximation \cite{Feyereisen/Fitzgerald/Komornicki:1993,Vahtras/Almlof/Feyereisen:1993,Weigend/Haser/Patzelt/Ahlrichs:1998}, 
which expands the products of molecular orbitals (MO) in terms of a set of auxiliary basis functions (ABFs).  
Our RI-based $GW$ implementation, together with our special on-the-fly procedure for constructing the ABFs, turns out to be remarkably accurate, as demonstrated in 
the $GW$100 project \cite{Setten/etal:2015}. 

In this work, we extend our molecular \GnWn implementation to periodic systems.
However, to extend a molecular $GW$ implementation to periodic systems within the NAO framework, several numerical obstacles have to be overcome. 
Within the family of local orbitals, we are aware of only a few periodic $GW$ implementations based on the LMTO method \cite{Aryasetiawan/Gunnarsson:1995} or on 
gaussian-type orbitals (GTOs) with the pseudopotential treatment of core ions \cite{Rohlfing/Kruger/Pollmann:1993,Wilhelm/Hutter:2017,Wilhelm:2018}, 
but to our knowledge there has been no reported NAO-based all-electron periodic $GW$ implementation. In the case of
ground-state KS-DFT,  the numerical techniques for periodic implementations within the NAO framework have been developed over the years and is now well established, as can be seen by the availability of a number of NAO-based DFT codes \cite{Delley:2000,Koepernik/Eschrig:1999,Soler/etal:2002,Ozaki/etal:2008,Blum/etal:2009,Kenney/Horsfield:2009,Li/Liu/etal:2016}. When coming to correlated methods, which require unoccupied states and two-electron integrals, NAO-based implementations
for periodic systems are still in their infancy. In the latter case, the major difficulties lie in the computation and storage of a 
large number of two-electron Coulomb repulsion integrals. Further challenges include how to construct efficient and high-quality NAO basis sets to describe the unoccupied energy states and how to treat the so-called Coulomb singularity 
in the Brillouin zone (BZ) sampling. In this paper, we will describe 
the algorithms and numerical techniques used in our periodic {\GnWn} implementation, focusing on our strategies to deal with 
the aforementioned challenging issues. Systematic convergence tests with respect to the computational parameters, and benchmarks against independent {\GnWn} implementations and experimental values for prototypical 3-dimensional (3D) semiconductors and insulators will be reported as well. Although not discussed here, we would like to 
mention that our {\GnWn} implementation can be readily applied to 1-dimensional (1D) and 2-dimensional (2D) systems, because our basis functions are strictly 
localized in space and as detailed in Sec.~\ref{sec:gamma}, the Coulomb operator is truncated at the boundary of the
supercell under the the Born-von-K\'{a}rm\'{a}n (BvK) periodic boundary condition. 
Furthermore, our implementation is highly parallel, scaling up to tens of thousands of CPU cores. In this paper, we focus on the basic algorithms and numerical precision
aspects, validating our implementation for 3D insulating systems. A discussion of the scalability and efficiency of implementation, as well as its performance
for low-dimensional systems, will be presented in a forthcoming paper.

This paper is organized as follows. In Sec.~\ref{sec:theory} we present the basic theory and algorithms behind our implementation, especially the working
equations for periodic $GW$ based on a localized variant of the RI approximation. In Sec.~\ref{sec:comput} the computational details will be discussed,
including our procedure to generate the one-electron orbital basis functions and auxiliary basis functions, and our algorithm to treat the Coulomb singularity. Systematic convergence tests with respect to
computational parameters are then presented in Sec.~\ref{sec:convergence} for Si and MgO. This is followed by Sec.~\ref{sec:benchmark} where benchmark 
calculations for a set of prototypical semiconductors and insulators are performed, and compared to well-established reference values. 
Finally we conclude our paper with an outlook in Sec.~\ref{sec:conclusion}.

\section{\label{sec:theory} Theoretical framework}
\subsection{$GW$ equations in real space}
In this section, we recapitulate the basic equations of the commonly used {\GnWn} approach, where both the Green function $G_0$ 
and screened Coulomb interaction $W_0$ are determined by the orbitals and orbital energies, obtained from a preceding (g)KS-DFT calculation. 
In real space, the \GnWn self-energy is given by
  \begin{widetext}
     \begin{equation}
	   \Sigma^{G_\text{0}W_\text{0}}_\sigma(\bfr,\bfrp,i\omega) = -\frac{1}{2\pi}\int_{-\infty}^\infty d\omegap G_{\text{0},\sigma}(\bfr,\bfrp,i\omega-i\omegap)
        W_\text{0}(\bfr,\bfrp,i\omegap)  
       \label{Eq:gw_selfenergy_realspace}
   \end{equation}
  \end{widetext}
where the non-interacting Green function $G_0$ is
   \begin{equation}
    G_{0,\sigma}(\bfr,\bfrp,i\omega) = \sum_{n,\bfk} w_\bfk \frac{\psi_{n,\sigma}^{\bfk}(\bfr)\psi^{\bfk\ast}_{n,\sigma}(\bfrp)}
	   {i\omega+\mu-\epsilon_{n,\sigma}^{\bfk}}\, .
    \label{Eq:green_function_realspace}
   \end{equation}
Here $\psi_{n,\sigma}^{\bfk}$ and $\epsilon_{n,\sigma}^{\bfk}$ are KS orbitals and orbital energies, with $n,\sigma$ being the orbital 
and spin indices, and $\bfk$ being a Bloch momentum vector in the 1st Brillouin zone (BZ). Furthermore,
$w_\bfk$ is integration weight of
the $\bfk$ point (for even-spaced $\bfk$ grids, $w_\bfk=1/N_\bfk$ with $N_\bfk$ being the number of $\bfk$ points in the 1BZ),
$\omega$ is a frequency point on the imaginary frequency axis, and $\mu$ is the electronic chemical potential.
The screened Coulomb interaction $W_0$ in Eq.~(\ref{Eq:gw_selfenergy_realspace}) is defined as
   \begin{equation}
     W_0(\bfr,\bfrp,i\omega)= \int d\bfrpp \varepsilon^{-1}(\bfr,\bfrpp,i\omega)v(\bfrpp,\bfrp)\, ,
   \end{equation}
where $\varepsilon^{-1}$ is the inverse of the microscopic dielectric function $\varepsilon$.  
 Within the random phase approximation (RPA), $\varepsilon$ is fully determined by the independent particle response function 
 $\chi_0$ and the bare Coulomb interaction $v$,
   \begin{equation}
      \varepsilon(\bfr,\bfrp,i\omega) = \delta(\bfr-\bfrp) - \int d\bfrpp v(\bfr,\bfrpp) \chi_{0}(\bfrpp,\bfrp,i\omega) \, .
   \end{equation}
The non-interacting response function $\chi_0$ can be expressed explicitly in terms of KS orbitals and 
orbital energies, according to the Adler-Wiser formula \cite{Adler:1962,Wiser:1963}
\begin{widetext}
\begin{equation}
   \chi_0(\bfr, \bfrp, i\omega) = 
   \sum_{m,n,\sigma}\sum_{\bfk,\bfq}^{\rm 1BZ} w_\bfk w_\bfq \frac{(f_{m,\sigma}^{\bfk+\bfq} - f_{n,\sigma}^{\bfq})
                   \psi^{\bfk+\bfq~\ast}_{m,\sigma}(\bfr)\psi_{n,\sigma}^{\bfk}(\bfr)
                    \psi^{\bfk\ast}_{n,\sigma}(\bfrp)\psi_{m,\sigma}^{\bfk+\bfq}(\bfrp) } {\epsilon_{m,\sigma}^{\bfk+\bfq}-
                    \epsilon_{n,\sigma}^{\bfk} -  i\omega} \, .
   \label{Eq:chi_0_realspace}
\end{equation}
\end{widetext}
In Eq.~(\ref{Eq:chi_0_realspace}) the summations of the Bloch vectors $\bfk, \bfq$ are over the 1BZ, 
and $f_{n,\sigma}^{\bfk}$ are the orbitals' occupation factors. In our formulation and  practical implementation to be described below, 
we work on the imaginary frequency axis. To get quasiparticle excitation energies, an analytical continuation of the self-energy from
	the imaginary to the real frequency axis will be carried out \cite{Rojas/Godby/Needs:1995}. Alternatively,
	the \GnWn self-energy on the real frequency axis can also be directly computed via the contour deformation (CD) approach.
	For molecules and clusers this has already been implemented in FHI-aims \cite{Golze/etal:2018}. It will be extended to the periodic case in the future. 

\subsection{Auxiliary basis representation of $GW$ equations} 
In practical calculations, $\chi_0(\bfr, \bfrp, i\omega)$ (as well as $\varepsilon$ and $W$) has to be discretized either on a 
real-space grid \cite{Rojas/Godby/Needs:1995} or more often it is expanded in terms of a suitable basis set. In the latter case,
	the basic choice to expand non-local quantities ($\chi_0$, $\varepsilon$, and $W_0$) depends on 
	the preceding computational framework to obtain the single-particle KS orbitals $\psi_{n,\sigma}^\bfk$. For example, 
	in the pseudopotential plane-wave \cite{Hybertsen/Louie:1986} or PAW frameworks \cite{Shishkin/Kresse:2006,Hueser/Olsen/Thygesen:2013}, these basis functions are simply plane waves. And in the LMTO
	\cite{Aryasetiawan/Gunnarsson:1994a} or LAPW \cite{Friedrich/etal:2009,Jiang/etal:2013,Jiang/Blaha:2016,Nabok/Gulans/Draxl:2016} frameworks, the so-called mixed product basis is used. Our own implementation employs the NAO basis function framework, whereby a set of atom-centered auxiliary 
	basis functions is constructed to expand the products of two KS orbitals. In the molecular case, this can be expressed as
    \begin{equation}
         \psi^\ast_{m,\sigma}(\bfr)\psi_{n,\sigma}(\bfr) \approx \sum_\mu C_{m,n,\sigma}^\mu P_\mu(\bfr)\,
         \label{eq:RI_expan_mol}
    \end{equation}
where $\psi_{n,\sigma}(\bfr)$ is a molecular KS orbital, $P_\mu(\bfr)$ is the $\mu$-th auxiliary basis function, 
 and $C_{m,n,\sigma}^\mu$ is the 3-orbital (triple) expansion coefficient. It follows that 
$\chi_0$, $\varepsilon$, and $W$ can all be represented in terms of $P_\mu(\bfr)$'s in a matrix form. These $P_\mu(\bfr)$ basis functions are termed ABFs, which are distinct from the orbital basis sets (OBS) $\{\varphi_i\}$ 
to expand a single KS orbital,
    \begin{equation}
      \psi_{n,\sigma}(\bfr) = \sum_i c_{i,n,\sigma} \varphi_i(\bfr-{\bm \tau}_i)\, .
	    \label{eq:ao2mo}
    \end{equation} 
  Here, $c_{i,n,\sigma}$ are the KS eigenvectors, and $\varphi_i$ is an orbital basis function (i.e., the NAO basis function 
  mentioned above) centered at the atomic position ${\bm \tau}_i$. Throughout this paper we use $m,n$ indices
for denoting KS orbitals, $i,j,k,l$ for atomic basic functions, and $\mu,\nu,\alpha,\beta$ for ABFs.

Our ABFs are also atom-centered, and they are constructed in a way similar to the mixed product basis in the LAPW framework, but without the plane-wave component in the interstitial region. 
In Ref.~\cite{Ren/etal:2012}, we have described how to make use of the expansion in Eq.~(\ref{eq:RI_expan_mol}) to achieve efficient
implementations of Hartree-Fock, 2nd-order M{\o}ller-Plesset perturbation theory (MP2), the random-phase approximation (RPA), and $GW$ within the NAO
	basis set framework. Approximations based on Eq.~(\ref{eq:RI_expan_mol}) are known as density fitting \cite{Whitten:1973,Dunlap/Connolly/Sabin:1979}, 
	or RI \cite{Feyereisen/Fitzgerald/Komornicki:1993,Vahtras/Almlof/Feyereisen:1993,Weigend/Haser/Patzelt/Ahlrichs:1998}
	in the context of evaluating two-electron Coulomb repulsion integrals, as already mentioned above. The RI-based implementation reported in Ref.~\cite{Ren/etal:2012}, 
though quite accurate for both Hartree-Fock and correlated methods, is restricted to molecular geometries under non-periodic boundary conditions. 
	More recently, we have extended our formalism and implementation to periodic systems. A 
periodic, linear-scaling Hartree-Fock and screened exact exchange implementation was reported in Ref.~\cite{Levchenko/etal:2015}. In this paper, 
we focus on the extension of our $GW$ implementation to periodic systems, based on the local variant of the RI approximation used in Refs.~\cite{Ihrig/etal:2015} and \cite{Levchenko/etal:2015}. 
     
In periodic systems, the KS orbitals carry an additional $\bfk$ vector in their indices, and their products can be expanded in terms of
the Bloch summation of the atom-centered ABFs,
 \begin{equation}
   \psi_{m,\sigma}^{\bfk+\bfq\ast}(\bfr)\psi_{n,\sigma}^{\bfk}(\bfr)=\sum_{\mu}^{N_\text{aux}} C_{m,n,\sigma}^\mu (\bfk+\bfq,\bfk) P_{\mu}^{\bfq\ast}(\bfr)\,
   \label{Eq:KSproduct_expan}
  \end{equation}
  where $N_\text{aux}$ is the number of ABFs within each unit cell,
  \begin{equation}
	  P_{\mu}^{\bfq}(\bfr) = \sum_{\bfR} P_{\mu}(\bfr-\bfR -{\bm \tau}_\mu)e^{i\bfq \cdot \bfR}\, ,
          \label{eq:ABF_FT}
  \end{equation}
  and $C_{m,n,\sigma}^\mu (\bfk+\bfq, \bfk)$ are the expansion coefficients which now depend on two independent Bloch wave vectors. In Eq.~(\ref{eq:ABF_FT}), ${\bm \tau}_\mu$ is
  the position of atom from which the $\mu$-th ABF originates within the unit cell, and the sum runs over all
  unit cells $\bfR$ in the BvK supercell.
 In our implementation, the atom-centered ABFs are chosen to be real-valued, and hence $P_{\mu}^{\bfq\ast}(\bfr)=P_{\mu}^{-\bfq}(\bfr)$.
 Using Eqs.~(\ref{Eq:chi_0_realspace}) and (\ref{Eq:KSproduct_expan}), one immediately arrives at
  \begin{equation}
    \chi_0(\bfr, \bfrp, i\omega) \approx \sum_{\mu,\nu} \sum_{\bfq} w_\bfk
	  P_{\mu}^{\bfq\ast}(\bfr) \chi_{0,\mu\nu}(\bfq,i\omega) P_{\nu}^{\bfq}(\bfrp) \, ,
   \label{Eq:chi_expan}
  \end{equation}
where the matrix representation of $\chi_0$ is given by
  \begin{equation}
     \chi_{0,\mu\nu}(\bfq, i\omega) = \sum_{\sigma,m,n} \sum_{\bfk} w_\bfk
       \frac{C_{m,n,\sigma}^\mu (\bfk+\bfq, \bfk) C_{n,m,\sigma}^\nu (\bfk, \bfk+\bfq) } 
        {\epsilon_{m,\sigma}^{\bfk+\bfq}-\epsilon_{n,\sigma}^{\bfk} - i\omega} \, .
   \label{Eq:chi0_matrix}
  \end{equation}

To obtain the matrix representation of $\varepsilon$ and $W_0$ in terms of ABFs, one still needs to compute
the Coulomb matrix given by expanding the Coulomb operator in terms of the same set of ABFs, 
   \begin{equation}
	   V_{\mu\nu}(\bfq) = \int d\bfr d\bfrp \frac{P_{\mu}^{\bfq\ast}(\bfr)P_{\nu}^{\bfq}(\bfrp)}{|\bfr-\bfrp|}   \, .
        \label{Eq:bare_V_matr}
   \end{equation}
   The matrix form of $\varepsilon$ and $W_0$ can then be obtained via matrix multiplication and inversion at each $(\bfk, i\omega)$ 
   point. For computational convenience, we use the symmetrized dielectric function 
   $\tilde{\varepsilon}=v^{-1/2}\varepsilon v^{1/2}$, whose matrix form can be computed as, 
   \begin{equation}
     \tilde{\varepsilon}_{\mu\nu}(\bfq,i\omega)=\delta_{\mu,\nu}-\sum_{\alpha\beta}
                          V_{\mu\alpha}^{1/2}(\bfq)\chi_{0,\alpha\beta}(\bfq,i\omega)V_{\beta\nu}^{1/2}(\bfq) \, 
    \label{Eq:symm_dielec_func}
   \end{equation}
   where $V^{1/2}$ is the square root of the $V$ matrix. The $\tilde{\varepsilon}$ matrix is then inverted and the matrix form of $W_0$ 
   can be obtained as
   \begin{equation}
       W_{0,\mu\nu}(\bfq,i\omega)=\sum_{\alpha,\beta}V^{1/2}_{\mu\alpha}(\bfq)\tilde{\varepsilon}^{-1}_{\alpha\beta}(\bfq,i\omega)
       V^{1/2}_{\beta\nu}(\bfq)
      \label{Eq:screened_W_matr}
   \end{equation}
   Noting that 
      \begin{equation}
        W_{0,\mu\nu}(\bfq,i\omega)=\iint d\bfr d\bfrp P_\mu^{\bfq \ast}(\bfr) W_0(\bfr,\bfrp,i\omega)P_\nu^{\bfq}(\bfrp)
      \end{equation}
      and using Eqs.~(\ref{Eq:gw_selfenergy_realspace}), (\ref{Eq:green_function_realspace}), and (\ref{Eq:KSproduct_expan}),
      one arrives at the following expression for computing the diagonal matrix element of the {\GnWn} self-energy,
      \begin{widetext}
       \begin{align}
         \Sigma_{n,\sigma}^{\GnWn}(\bfk,i\omega) &= 
                        \iint d\bfr d\bfrp \psi_{n,\sigma}^{\bfk\ast}(\bfr) \Sigma^{G_\text{0}W_\text{0}}_\sigma(\bfr,\bfrp,i\omega) 
                    \psi_{n,\sigma}^{\bfk}(\bfrp)  \\
                   & = - \frac{1}{2\pi} \sum_{m,\bfq}\sum_{\mu,\nu}
            \int_{-\infty}^\infty d\omegap \frac{C_{n,m,\sigma}^\mu(\bfk,\bfk-\bfq)W_{0,\mu\nu}(\bfq,i\omegap)C_{m,n,\sigma}^\nu(\bfk-\bfq,\bfk)}
          {i\omega-i\omegap + \mu - \epsilon_{m,\sigma}^{\bfk-\bfq}}\, .
         \label{Eq:GW_selfe_element}
       \end{align}
      \end{widetext}
  In this formulation, the equations are well defined and the quantities can in principle be evaluated straightforwardly except at
  the $\Gamma$ ($\bfq=0$) point where elements of the $V$ and $W$ matrices between two nodeless $s$ functions, or between one nodeless $s$ and one
  nodeless $p$ function will diverge. Simply avoiding the $\Gamma$ point in the
  BZ sampling is not an optimal solution since this can result in a prohibitively slow convergence with respect to the summation over the $\bfk$ points.
  Because of the localized, non-orthogonal nature of our ABFs, the $\Gamma$-point correction schemes developed in the context of plane-wave \cite{Baroni/Resta:1986,Hybertsen/Louie:1987} or LAPW basis set framework \cite{Friedrich/etal:2009} are not directly applicable here. We will discuss our treatment of the $\Gamma$-point singularity in Sec.~\ref{sec:gamma}.
  
\subsection{\label{sec:gw_lri}Localized resolution of identity technique to determine the expansion coefficients}
Obviously, in the above formulation, the key issues are: $i)$ To construct a proper, sufficiently accurate 
auxiliary basis set $\{P^\bfq_\mu(\bfr)\}$, and $ii)$ to determine the expansion coefficients 
$C_{m,n,\sigma}^\mu(\bfk+\bfq,\bfk)$ which are needed in the computation of both $\chi_0$ matrix in Eq.~(\ref{Eq:chi0_matrix}) 
and {\GnWn} self-energy element in Eq.~(\ref{Eq:GW_selfe_element}). Our procedure to construct
the ABFs and the precision that can be achieved in practical calculations have been discussed in previous works 
\cite{Ren/etal:2012,Ihrig/etal:2015,Levchenko/etal:2015}. We will come to this point only briefly in Sec.~\ref{sec:basis},
 in the context of periodic {\GnWn} calculations. In this subsection, we will focus on the point $ii)$ and discuss
 how the expansion coefficients $C_{m,n,\sigma}^\mu(\bfk+\bfq,\bfk)$ are determined.   

In FHI-aims \cite{Blum/etal:2009} the KS orbitals are expanded in terms of NAO basis functions,
  \begin{align}
	  \psi_{n\sigma}^{\bfk}(\bfr) & = \sum_{i=1}^{N_\textnormal{b}} c_{i,n,\sigma}(\bfk)\varphi_{i}^{\bfk}(\bfr)  \nonumber \\
	  & = \sum_{i=1}^{N_\text{b}} c_{i,n,\sigma}(\bfk) \sum_{\bfR} \varphi_{i}(\bfr-\bfR-{\bm \tau}_i) e^{i\bfk\cdot \bfR} 
   \label{Eq:KS_orbital}
  \end{align}
where $c_{i,n,\sigma}(\bfk)$ are the KS eigenvectors, $N_\text{b}$ is the number of basis functions within one
unit cell, $\bfR$ is a Bravais lattice vector, and ${\bm \tau}_i$ is the position of the atom (within the unit cell) on which the basis function
$i$ is centered. In Eq.~(\ref{Eq:KS_orbital}), the atomic function $\varphi_i$ is given by
  \begin{equation}
    \varphi_i(\bfr)=u_{a,s,l}(r)Y_{lm}(\hat{\bfr}) \, .
    \label{Eq:basis_func}
  \end{equation}
  where $u_{a,s,l}(r)$ is the radial function, and $Y_{lm}(\hat{\bfr})$ is a spherical harmonic [in our implementation real-valued harmonic, i.e.,
  the real (cosine, positive $m$ values) or imaginary (sine, negative $m$ values) component of a spherical harmonic]. 
Thus the atomic orbital $i$ is fully specified by the atom index $a$, the radial function index $s$, and the angular momentum indices $l,m$.  

The ABFs used in FHI-aims are also atom-centered, but with different radial functions,
\begin{equation}
  P_\mu(\bfr)=\xi_{a,s,l}(r)Y_{lm}(\hat{\bfr}),
\end{equation}
where $\xi_{a,s,l}(r)$ is the radial auxiliary function, and $\mu$ is also a combined basis index of $a,s,l$, and $m$. 
The radial auxiliary functions are constructed in a way \cite{Ren/etal:2012,Ihrig/etal:2015} that the products of the KS orbitals 
$\psi_{m\sigma}\psi_{n\sigma}$, or equivalently the products of orbital basis functions $\varphi_i\varphi_j$ can be 
represented by a linear combination of $\{P_\mu(\bfr)\}$.  


To compute the expansion coefficients $C_{m,n,\sigma}^{\mu}(\bfk+\bfq,\bfk)$, defined in Eq.~(\ref{Eq:KSproduct_expan}), we 
first determine the expansion coefficients of the products of two (Bloch summed) basis functions 
$\varphi_i^{\bfk+\bfq \ast}(\bfr)\varphi_j^{\bfk}(\bfr)$ in terms of the ABFs,
  \begin{equation}
	  \varphi_{i}^{\bfk+\bfq\ast}(\bfr)\varphi_{j}^{\bfk}(\bfr) = \sum_{\mu=1}^{N_\text{aux}} \tilde{C}_{i,j}^\mu(\bfk+\bfq, \bfk) 
	   P_{\mu}^{\bfq\ast}(\bfr)\, ,
    \label{Eq:Bloch_NAOproduct_expan}
  \end{equation}
  and then transform $\tilde{C}_{i,j}^\mu(\bfk+\bfq, \bfk)$ to the $C_{m,n,\sigma}^{\mu}(\bfk+\bfq,\bfk)$ by multiplying with the KS eigenvectors,
  \begin{equation}
        C_{m,n,\sigma}^{\mu}(\bfk+\bfq,\bfk) = \sum_{i,j} c_{i,m,\sigma}^\ast(\bfk+\bfq)c_{j,n,\sigma}(\bfk)
	  \tilde{C}_{i,j,\sigma}^\mu(\bfk+\bfq,\bfk)\, .
       \label{Eq:AO_MO_transform}
  \end{equation}
  Below we shall refer to $C_{m,n,\sigma}^{\mu}(\bfk+\bfq,\bfk)$ as molecular orbital (MO) triple coefficients
  and $\tilde{C}_{i,j,\sigma}^\mu(\bfk+\bfq,\bfk)$ as atomic orbital (AO) triple coefficients.
Both types of triple coefficients depend on three orbital indices and in addition on two independent momentum vectors
$\bfk$ and $\bfq$. Take the AO triple coefficients for example, the number of entries scales as $N_\text{b}*(N_\textnormal{b}+1)/2*N_\text{aux}*N_{\bfk}^2$, 
and it is quite expensive to compute and store them. In FHI-aims
we can adopt the LRI approximation \cite{Ihrig/etal:2015} 
to deal with this issue. Within the LRI approximation, the ABFs used to
expand the product of two NAOs are restricted to those centering on the two atoms on which these two NAOs are centered. In quantum chemistry, such a two-center LRI scheme is also known as pair-atom RI (PARI) approximation \cite{Merlot/etal:2013,Wirz/etal:2017}. In real space, the two NAOs $i$, $j$ in general can originate from
two different unit cells, labeled by two Bravais lattice vectors $\bfR_i$ and $\bfR_j$. The LRI approximation for periodic systems then implies that
  \begin{widetext}
   \begin{equation}
      \varphi_i(\bfr-\bfR_i-{\bm \tau}_i)\varphi_j(\bfr-\bfR_j-{\bm {\bm \tau}}_j) \approx \sum_{\mu \in I} 
	    \tilde{C}_{i(\bfR_i),j(\bfR_j)}^{\mu(\bfR_i)} P_\mu(\bfr-\bfR_i-{\bm {\bm \tau}}_i) + 
	    \sum_{\mu \in J} \tilde{C}_{i(\bfR_i),j(\bfR_j)}^{\mu(\bfR_j)} P_\mu(\bfr-\bfR_j-{\bm {\bm \tau}}_j)\, ,
      \label{Eq:realspace_expan1}
   \end{equation}
  \end{widetext}
where $\tilde{C}_{i(\bfR_i),j(\bfR_j)}^{\mu(\bfR_i)}$ are the two-center expansion coefficients 
where the lattice vector in parenthesis associated with the basis index indicates the unit cell from which the basis function
originates. Furthermore, $I,J$ in Eq.~(\ref{Eq:realspace_expan1}) denote the atoms where $\varphi_i$ and $\varphi_j$ are centered, and
$\mu \in I$ means the summation over the ABFs is restricted to those centering at the atom $I$.  Because of the translational symmetry of 
the periodic system, one has 
      $C_{i(\bfR_i),j(\bfR_j)}^{\mu(\bfR_i)}=C_{i(\bfzero),j({\bfR}_j-{\bfR}_i)}^{\mu(\bfzero)}$, where $\bfzero$ here denotes
the unit cell at the origin. Therefore Eq.~(\ref{Eq:realspace_expan1}) becomes
   \begin{widetext}
	   \begin{equation}
	      \varphi_i(\bfr-\bfR_i-{\bm  \tau}_i)\varphi_j(\bfr-\bfR_j-{\bm  \tau}_j) \approx \sum_{\mu \in I}  
		    \tilde{C}_{i(\bfzero),j(\bfR_j-\bfR_i)}^{\mu(\bfzero)} P_\mu(\bfr-\bfR_i-{\bm \tau}_i) +  \sum_{\mu \in J}
		    \tilde{C}_{i(\bfR_i-\bfR_j),j(\bfzero)}^{\mu(\bfzero)} P_\mu(\bfr-\bfR_j-{\bm \tau}_j)\, ,
	      \label{Eq:realspace_expan2}
	   \end{equation}
   \end{widetext}
 This means that the two-center expansion coefficients in real space naturally split into two sectors, and each of them only 
 depends on one independent lattice vector. Now, by Fourier transforming Eq.~(\ref{Eq:realspace_expan2}) to $\bfk$ space from both sides, we obtain
 \begin{widetext}
 \begin{align}
	 & \varphi_{i}^{\bfk+\bfq\ast}(\bfr) \varphi_{j}^{\bfk}(\bfr) = \sum_{\bfR_i,\bfR_j} e^{-i(\bfk+\bfq)\cdot\bfR_i} e^{i\bfk\cdot\bfR_j}
	   \varphi_{i}(\bfr-\bfR_i -{\bm \tau}_i)\varphi_{j}(\bfr-\bfR_j - {\bm \tau}_j) \nonumber \\
        \approx & \sum_{\bfR_i,\bfR_j} e^{-i(\bfk+\bfq)\cdot\bfR_i} e^{i\bfk\cdot\bfR_j}
		  \left[ \sum_{\mu \in I} \tilde{C}_{i(\bfzero),j(\bfR_j-\bfR_i)}^{\mu(\bfzero)}
		  P_{\mu}(\bfr-\bfR_i - {\bm \tau}_i) + \sum_{\mu \in J} 
		    \tilde{C}_{i(\bfR_i-\bfR_j), j(\bfzero)}^{\mu(\bfzero)} P_{\mu}(\bfr-\bfR_j-{\bm \tau}_j) \right] \nonumber  \\
	    = & \sum_{\mu \in I} \left[ \sum_{\bfR_i} e^{-i\bfq\cdot\bfR_i}P_{\mu}(\bfr-\bfR_i -{\bm \tau}_i)\sum_{\bfR_j}e^{i\bfk\cdot(\bfR_j-\bfR_i)} 
		 \tilde{C}_{i(\bfzero),j(\bfR_j-\bfR_i)}^{\mu(\bfzero)} \right] +  \nonumber \\
	      	&    \sum_{\mu \in J} \left[ \sum_{\bfR_j} e^{-i\bfq\cdot\bfR_j}P_{\mu}(\bfr-\bfR_j-{\bm \tau}_j)
		   \sum_{\bfR_i}e^{-i(\bfk+\bfq)\cdot(\bfR_i-\bfR_j)} \tilde{C}_{i(\bfR_i-\bfR_j),j(\bfzero)}^{\mu(\bfzero)} \right] \nonumber \\
	   = & \sum_{\mu \in I} \tilde{C}_{i(-\bfk-\bfq),j(\bfzero)}^{\mu(\bfzero)} P_\mu^{\bfq\ast}(\bfr) + 
	    \sum_{\mu \in J}\tilde{C}_{i(\bfzero),j(\bfk)}^{\mu(\bfzero)} P_\mu^{\bfq\ast}(\bfr) \, .
	   \label{Eq:local_ri_FT}
 \end{align}
 \end{widetext}
 Here we have introduced the notation
    \begin{align}
	    \tilde{C}_{i(\bfk),j(\bfzero)}^{\mu(\bfzero)} & = \sum_{\bfR} e^{i\bfk \cdot \bfR} \tilde{C}_{i(\bfR),j(\bfzero)}^{\mu(\bfzero)}\, \label{Eq:expan_coeff_FT1} ,    \\
	    \tilde{C}_{i(\bfzero),j(\bfk)}^{\mu(\bfzero)} & = \sum_{\bfR} e^{i\bfk \cdot \bfR} \tilde{C}_{i(\bfzero),j(\bfR)}^{\mu(\bfzero)}\, . 
	    \label{Eq:expan_coeff_FT2}
    \end{align}
To derive the last line of Eq.~(\ref{Eq:local_ri_FT}), we have used Eq.~(\ref{eq:ABF_FT}), and realize that ${\bm \tau}_\mu = {\bm \tau}_i$ in the first term and  
${\bm \tau}_\mu={\bm \tau}_j$ in the second term.

 Comparing Eq.~(\ref{Eq:local_ri_FT}) to (\ref{Eq:Bloch_NAOproduct_expan}), we arrived at the following desired result,
  \begin{equation}
	  \tilde{C}_{i,j}^\mu(\bfk+\bfq, \bfk) =\left\{ \begin{array}{lcl} \tilde{C}_{i(-\bfk-\bfq),j(\bfzero)}^{\mu(\bfzero)} &, &  ~~ \mu \in I \\
		  \tilde{C}_{i(\bfzero),j(\bfk)}^{\mu(\bfzero)} & , & ~~ \mu \in J  \\
									  0 &, & ~~ \textnormal{otherwise}  \\
                                                        \end{array}
                                                \right.
	  \label{Eq:AO_triple_split}
  \end{equation}
  Equation~(\ref{Eq:AO_triple_split}) indicates that AO triple coefficients have only two non-zero sectors, each of which only depends on one independent $\bfk$ 
  vector instead of two. This property greatly simplifies the computation and storage of the triple coefficients, which are the key quantities in 
   our {\GnWn} implementation. 

  The evaluation of two-center expansion coefficients $\tilde{C}_{i(\bfzero),j(\bfR)}^{\mu(\bfzero)}$ is described in detail in Ref.~\cite{Ihrig/etal:2015}. 
  Essentially, they are determined by minimizing the self Coulomb repulsion of the expansion error given by Eq.~(\ref{Eq:realspace_expan2}). This criterion leads
  to the following expression, 
    \begin{equation}
	    \tilde{C}_{i(\bfzero),j(\bfR)}^{\mu(\bfzero)} =\left\{\begin{array}{c}\displaystyle \sum_{\nu\in \{I,J(\bfR)\}} (i(\bfzero),j(\bfR)|\nu) \left(V^{IJ}\right)^{-1}_{\nu\mu}, ~\text{for}~ \mu \in I \\
		    0, ~~~\text{otherwise}
	    \end{array} \right.
	    \label{Eq:two_center_coeff_determination}
    \end{equation}
    where $\nu\in \{I,J(\bfR)\}$ means that the auxiliary function
    $P_\nu$ is centered either on the atom $I$ in the original cell, or on the atom $J$ in the cell specified by $\bfR$. Furthermore $(i,j|\nu)$ is the Coulomb
    repulsion between the product $\varphi_i\varphi_j$ and the ABF $P_\mu$,
      \begin{equation}
	      (i,j|\mu) = \iint d\bfr d\bfrp \frac{\varphi_i(\bfr)\varphi_j(\bfr)P_\mu(\bfrp)}{|\bfr-\bfrp|} \, ,
      \end{equation}
      and $V^{IJ}$ is a sub-block of the Coulomb matrix $V$  where the auxiliary basis indices of the entries belong to either atom $I$ or atom $J$. 
      Only an inversion of such a sub-block of the Coulomb matrix is required at a time to determine the two-center expansion coefficients in
      Eq.~(\ref{Eq:two_center_coeff_determination}). 
      We note that, in Eq.~(\ref{Eq:two_center_coeff_determination}), only two-center integrals (albeit three orbitals are involved) are required, thanks
      to the LRI approximation. Efficient algorithms exist to evaluate two-center integrals over numeric atom-centered basis functions 
      \cite{Talman78-SBT,Talman84-4center,Talman03-MCI}.
      All these pieces add together to make an efficient evaluation of the AO triple coefficients possible.
      
      Furthermore, one may observe that $\tilde{C}_{i(\bfzero),j(\bfR)}^{\mu(\bfzero)}=\tilde{C}_{j(\bfR),i(\bfzero)}^{\mu(\bfzero)}$, and
      hence according to Eqs.~(\ref{Eq:expan_coeff_FT1}) and (\ref{Eq:expan_coeff_FT2}), we have
      $\tilde{C}_{j(\bfk),i(\bfzero)}^{\mu(\bfzero)}=\tilde{C}_{i(\bfzero),j(\bfk)}^{\mu(\bfzero)}$.  It follows that Eq.~(\ref{Eq:AO_triple_split}) 
     can be rewritten as
      \begin{equation}
	      \tilde{C}_{i,j}^\mu(\bfk+\bfq), \bfk) = \tilde{C}_{j(\bfzero),i(-\bfk-\bfq)}^{\mu(\bfzero)} +\tilde{C}_{i(\bfzero),j(\bfk)}^{\mu(\bfzero)}
	  \label{Eq:AO_triple_split2}
      \end{equation}

\begin{figure}
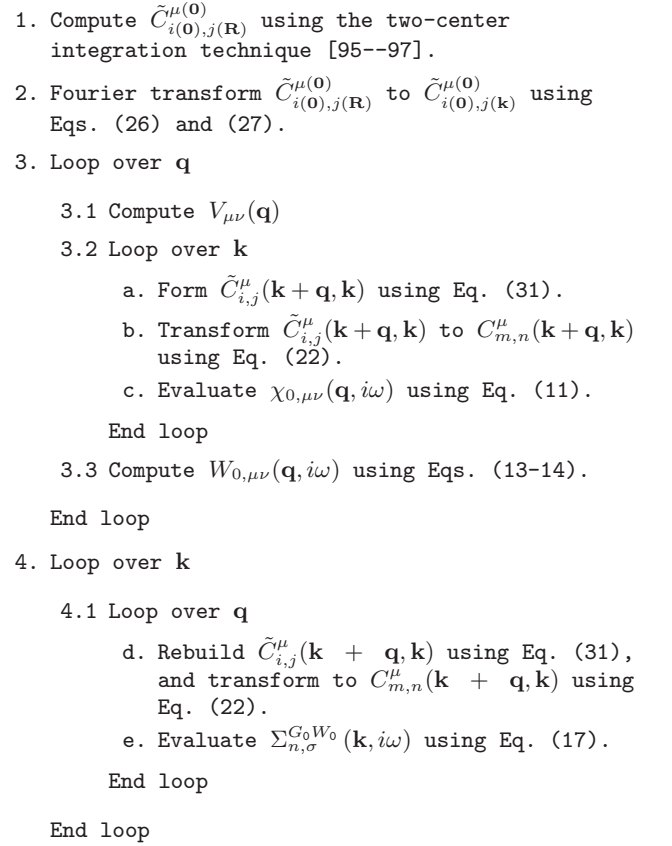

      \vskip 1mm
	  \texttt{
	   \rule{0.5\textwidth}{0.7pt}
	  \begin{enumerate}
		  \item Compute $\tilde{C}_{i(\bfzero),j(\bfR)}^{\mu(\bfzero)}$ using the two-center integration technique 
                        \cite{Talman78-SBT,Talman84-4center,Talman03-MCI}.
		  \item Fourier transform $\tilde{C}_{i(\bfzero),j(\bfR)}^{\mu(\bfzero)}$ to $\tilde{C}_{i(\bfzero),j(\bfk)}^{\mu(\bfzero)}$ using
                        Eqs.~(\ref{Eq:expan_coeff_FT1}) and (\ref{Eq:expan_coeff_FT2}).
		  \item Loop over $\bfq$ \\
         		  \begin{enumerate}
				  \item[3.1] Compute $V_{\mu\nu}(\bfq)$
				  \item[3.2]  Loop over $\bfk$ \\
			  \begin{enumerate}
				  \item[a.]  Form $\tilde{C}_{i,j}^\mu(\bfk+\bfq,\bfk)$ using Eq.~(\ref{Eq:AO_triple_split2}).
				  \item[b.] Transform $\tilde{C}_{i,j}^\mu(\bfk+\bfq,\bfk)$ to $C_{m,n}^{\mu}(\bfk+\bfq,\bfk)$ 
					    using Eq.~(\ref{Eq:AO_MO_transform}).
				  \item[c.] Evaluate $\chi_{0,\mu\nu}(\bfq,i\omega)$ using Eq.~(\ref{Eq:chi0_matrix}).
                          \end{enumerate}
		             \item[] End loop 
		             \item[3.3] Compute $W_{0,\mu\nu}(\bfq,i\omega)$ using Eqs.~(\ref{Eq:symm_dielec_func}-\ref{Eq:screened_W_matr}).
	               \end{enumerate}
	     \item[] End loop
	  \end{enumerate}
	  \begin{enumerate}
		  \item[4.] Loop over $\bfk$ \\
	           \begin{enumerate}
			   \item[4.1] Loop over $\bfq$ \\
			  \begin{enumerate}
				  \item[d.]  Rebuild $\tilde{C}_{i,j}^\mu(\bfk+\bfq,\bfk)$ using Eq.~(\ref{Eq:AO_triple_split2}), and
				            transform to $C_{m,n}^{\mu}(\bfk+\bfq,\bfk)$ using Eq.~(\ref{Eq:AO_MO_transform}).
				    \item[e.] Evaluate $\Sigma_{n,\sigma}^{\GnWn}(\bfk,i\omega)$ using Eq.~(\ref{Eq:GW_selfe_element}).
                          \end{enumerate}
		       \item[] End loop 
	           \end{enumerate}
	     \item[] End loop
	  \end{enumerate}
	  \vskip -3mm
	  \rule{0.5\textwidth}{0.7pt}
	  }
	  \caption{Flowchart of the $\GnWn$ self-energy calculation for periodic systems within the NAO framework and LRI approximation.}
	  \label{Fig:flowchart_G0W0}
  \end{figure}

      We summarize the computational algorithm described above in terms of the flowchart in Fig.~\ref{Fig:flowchart_G0W0}. The key in this algorithm is that
      we only need to explicitly store $\tilde{C}_{i(\bfzero),j(\bfR)}^{\mu(\bfzero)}$ and its Fourier transform $\tilde{C}_{i(\bfzero),j(\bfk)}^{\mu(\bfzero)}$.
      The memory intensive $\tilde{C}_{i,j}^\mu(\bfk+\bfq,\bfk)$ and $C_{m,n}^{\mu}(\bfk+\bfq,\bfk)$ are formed on the fly when needed, within the loop over
      the $\bfk$ and $\bfq$ points. This efficacy of this algorithm depends on the accuracy of the LRI approximation, which further depends on the 
      auxiliary basis set $\{P_{\mu}(\bfr)\}$. We will discuss this issue in the next section.

The algorithm as outlined in Fig.~\ref{Fig:flowchart_G0W0} scales as $O(N^4)$ with respect to the number of one-electron basis functions and quadratically with respect to the number of \bfk points. In the literature, a $O(N^3)$-scaling $GW$ algorithm has been formulated based on real-space/imaginary-time representation of the response function \cite{Rojas/Godby/Needs:1995,GW_space-time_method:1998}, and is getting popular in recent years as manifested in several recent implementations within different basis set frameworks \cite{Foerster/etal:2011,Kutepov/etal:2012,PeitaoLiu/etal:2016}.  Within the NAO framework, a similar real-space algorithm can be derived, which, fostered by the LRI approximation, features an $O(N^2)$ scaling of rate-determining steps of $GW$ self-energy calculations. The implementation work based on this algorithm is still going on. The present canonical-scaling implementation provides the reference results that one should be able to reproduce with a correct implementation of the aforementioned $O(N^2)$ real-space algorithm.


%

 \section{\label{sec:comput}Implementation details}
In the previous section, we presented the basic equations behind our implementation. Two important aspects that remain to be addressed are the procedure
to construct the ABFs, which is crucial for the accuracy of the LRI approximation and the treatment of the Coulomb singularity at the $\Gamma$ point.
We will focus on these two aspects in this section.
 \subsection{\label{sec:basis} Basis sets}
In FHI-aims, the basis functions to represent the KS orbitals are given by numerically tabulated radial functions multiplied by spherical harmonics,
as indicated by Eq.~(\ref{Eq:basis_func}). The radial functions $u_{a,s,l}(r)$ are usually constructed to satisfy the radial Schr\"{o}dinger equation 
(assuming the Hartree atomic unit, and abbreviating the indices $\{a,s,l\}$ by $j$),
  \begin{widetext}
   \begin{equation}
    \left( -\frac{1}{2r^2}\frac{d}{dr}r^2 \frac{d}{dr}  + \frac{l(l+1)}{r^2} + v_j(r) + v_\textnormal{conf}(r) \right)u_{j}(r) = \epsilon_{j} u_{j}(r),
     \label{Eq:radial_schrondinger}
   \end{equation}
  \end{widetext}
where $v_j(r)$ is a radial potential that defines the major behavior of $u_{j}(r)$, whereas $v_\textnormal{conf}(r)$ is a confining potential, 
added here in order to strictly localize $u_{j}(r)$ within a cutoff radius. The choice of $v_\textnormal{conf}(r)$ in FHI-aims is discussed in Ref.~\cite{Blum/etal:2009}.
The potential $v_j(r)$ is set to be the self-consistent free-atom potential for the atomic species $s$ in question, to obtain the so-called \textit{minimal}
basis, consisting of core and valence wave functions of a spherically symmetric atom. Additional basis functions beyond \textit{minimal} basis are of
ionic or hydrogen type, obtained by setting $v_j(r)$ to the potential of the cations of the atomic species $s$, or simply the hydrogen-like potential  
$Z/r$. The effective charge $Z$, being fractional in general, is taken as an optimization parameter, which controls the shape and spatial extension of of the hydrogen-like orbital basis functions.  These additional basis functions (mostly being
hydrogen-type) are selected by optimizing the ground-state total energy of target systems and grouped into different levels.
In FHI-aims, we have created  two types of such hierarchical NAO basis sets. One type is the FHI-aims-2009 basis sets (often called \textit{tier}-$n$ basis), 
originally optimized for ground-state DFT calculations for symmetric dimers of varying bond lengths \cite{Blum/etal:2009}, 
but proved to be also useful for {\GnWn} calculations 
\cite{Ren/etal:2012,Caruso/etal:2013,Liuchi/etal:2020}. Another type is the NAO-VCC-$n$Z ($n=2,3,4,5$) basis sets \cite{IgorZhang/etal:2013}, generated by optimizing the
RPA total energy of free atoms. The construction of the latter type of NAO basis sets follows the ``correlation consistent" (cc) strategy 
of Dunning \cite{Dunning:1989}, and hence allows for extrapolations and suitable for correlated calculations (such as MP2, RPA, and $GW$). 
In this work, we will check the performance of both types of basis sets for periodic $GW$ calculations.

The ABFs $\{P_\mu(\bfr)\}$ are not pre-constructed and optimized, but are rather formed adaptively based on a given OBS $\{\varphi_i(\bfr)\}$.
The detailed procedure to construct $\{P_\mu(\bfr)\}$ for a given set of $\{\varphi_i(\bfr)\}$ has been described in Refs.~\cite{Ren/etal:2012,Ihrig/etal:2015}.
The essential point is that the radial functions of $\{P_\mu(\bfr)\}$ are generated from the ``on-site" products of the radial functions of $\{\varphi_i(\bfr)\}$,
and the Gram-Schmidt procedure is then used to remove the linear dependence according to a threshold $\theta_\text{orth}$. The standard ABFs, generated from
the OBS used in the preceding self-consistent field (SCF) calculations, are sufficiently accurate for post-DFT correlated calculations, if used in the global Coulomb-metric
RI (called RI-V in Ref.~\cite{Ren/etal:2012}) framework. However, these ABFs alone are not adequate to yield the needed accuracy 
when used in the 
LRI scheme, especially for correlated methods that require unoccupied orbitals \cite{Ihrig/etal:2015}. The LRI approximation employed in the present work corresponds
to its non-robust fitting formulation, and the incurred error in the two-electron Coulomb integrals are linear with respect to the expansion error
of the orbital products [cf. Eq.~(\ref{eq:RI_expan_mol})] \cite{Merlot/etal:2013,Wirz/etal:2017}, in contrast with the RI-V case where the error is quadratic. One way to remedy this problem is to complement
the OBS with extra basis functions (called OBS+) that are used only for generating ABFs, but not in the preceding SCF calculations. It turns out that
this is a very efficient way to improve upon the standard ABF set, rendering the LRI a sufficiently accurate approximation for practical calculations.
In Ref.~\cite{Ihrig/etal:2015}, it was found that adding an extra $5g$ hydrogen-like function (with $Z=6$) to OBS results in an ABF set that is
sufficiently accurate for both ground-state exact-exchange and correlated calculations (MP2 and RPA), for the molecules/clusters tested in that work.
It should be noted though that the resulting ABFs for LRI can reach quite high angular momenta, e.g., up to $l=8$ if the OBS+ basis set includes angular
momenta up to $l=4$ (see Table 12 in Ref.~\cite{Ihrig/etal:2015} for an example). These high angular momenta turn out to be important for
the success of the LRI strategy.

In this work, we will test if the strategy of adding a $5g$ to OBS+ also works for periodic $GW$ calculations. As will be demonstrated in Sec.~\ref{sec:conv_aux},
for the $GW$ case a combined $4f5g$ OBS+ can yield excellent accuracy, and will be chosen as the default setting in production calculations.

\subsection{\label{sec:gamma} The $\Gamma$-point singularity treatment}

For 3D systems, the $1/r$ nature of the bare Coulomb potential leads to a $1/q^2$ divergence for $q\rightarrow 0$ in reciprocal space.
Within the plane wave basis, the Coulomb operator has a well-known matrix form $V_{\bfG,\bfGp}(\bfq)= 4\pi\delta_{\bfG,\bfGp}/|\bfq+\bfG|^2$, and the divergence is
only present in the $\bfG=\bfGp=0$ element (the so-called ``head" term of a matrix with indices $\bfG$ and $\bfGp$). With the atom-centered ABFs used in this work, 
this divergence carries over
to the matrix elements between two nodeless $s$-type functions ($1/q^2$ divergence), and between one nodeless $s$-type and one nodeless $p$-type functions 
($1/q$ divergence). In general
we can split $V_{\mu\nu}(\bfq)$ into
  \begin{equation}
    V_{\mu\nu}(\bfq) = \frac{v_{\mu\nu}^{(2)}}{q^2} + \frac{v_{\mu\nu}^{(1)}}{q} + \bar{V}_{\mu\nu}(\bfq)\, ,
    \label{Eq:bare_V_symptotic}
  \end{equation}
where $\bar{V}_{\mu\nu}(\bfq)$ is regular as $\bfq \rightarrow \bfzero$, and $v_{\mu\nu}^{(2)}$ and $v_{\mu\nu}^{(1)}$ are the coefficients for elements 
with $1/q^2$ and $1/q$ asymptotic behaviors. Simple electrostatic analysis indicates that $v_{\mu\nu}^{(2)}$ and $v_{\mu\nu}^{(1)}$ are only nonzero for 
the above-noted pair of basis functions that have $1/q^2$ and $1/q$ divergences respectively.

The above diverging behavior of $V_{\mu\nu}(\bfq)$ carries over to the screened Coulomb matrix $W_{0,\mu\nu}(\bfq,i\omega)$
for non-metallic systems. Consequently, in the integration over BZ to obtain the {\GnWn} self energy, the integrand 
has a $1/q^2$ 
divergence as $\bfq$ approaches the $\Gamma$ point. This is an integrable divergence. However, when the BZ is discretized in terms of
a uniform $\bfk$ mesh, the $\Gamma$ point contributes to the integrated quantity -- the {\GnWn} self energy -- by an amount 
that is proportional to the discretization length $\Delta q$, i.e., the length of the small cube enclosing the $\Gamma$ point. 
Since $\Delta q \sim N_\bfk^{-1/3}$, simply neglecting the $\Gamma$ point will incur a so-called ``finite-size error" 
that decreases only linearly with with $N_\bfk^{-1/3}$. This is a very slow convergence. 
Therefore, to achieve a sufficiently fast convergence of the BZ integration
in Eq.~(\ref{Eq:GW_selfe_element}), a special treatment of this singularity is needed.
For semiconductors and insulators, the (properly treated) dielectric function is non-diverging around $\bfq = \bfzero$, 
which means that $W_{0,\mu\nu}(\bfq,i\omega)$ 
has the same asymptotic behavior as $V_{\mu\nu}(\bfq)$ for $\bfq \rightarrow \bfzero$. 
Thus, the experience gained to treat the singularity of the bare Coulomb 
potential in periodic Hartree-Fock (HF) calculations is also useful here. 
In the literature, two schemes are widely adopted to deal with the Coulomb singularity 
in periodic HF implementations. One is the Gygi-Baldereschi scheme \cite{Gygi/Baldereschi:1986} 
which adds an analytically integrable compensating function to 
cancel the diverging term and subtracts it separately. The other is the Spencer-Alavi scheme \cite{Spencer/Alavi:2008} which uses a truncated Coulomb operator 
that is free of Coulomb singularity; yet the scheme guarantees a systematic convergence to the right limit 
as the number of $\bfk$ points increases. Both schemes have been implemented 
in our own periodic HF module \cite{Levchenko/etal:2015} of the FHI-aims code. In practice, we found that 
a modified version of the Spencer-Alavi scheme converges faster with respect to the number of $\bfk$ points than 
the ``compensating function" approach does. Similar observations have been made in Ref.~\cite{Sundararaman/Arias:2013}, 
in terms of another variant of the Coulomb operator truncation scheme -- the so-called Wigner-Seitz cell truncation scheme \cite{Sundararaman/Arias:2013}. 

In our modified Spencer-Alavi scheme, the truncated Coulomb operator is given by,
\begin{widetext}
     \begin{equation}
      v^{cut}(r) = \frac{\text{erfc}(\gamma r)}{r}  + 0.5*\text{erfc}\left[(\text{ln}(r)-\text{ln}(R_{cut}))/\text{ln}(R_{w})\right]\frac{\text{erf}(\gamma r)}{r}\, 
             \label{Eq:cut_coulomb_pot}
     \end{equation}
\end{widetext}
where the short-range part of the Coulomb potential is kept, and the long-range part is quickly suppressed beyond a cutoff radius $R_{cut}$. $R_{cut}$ 
is chosen to be the radius of a sphere inscribed inside the BvK supercell. As the $\bfk$ point mesh gets denser, $R_{cut}$ gradually 
increases and the full bare Coulomb operator is restored. The screening parameter $\gamma$ and the width parameter $R_{w}$ in Eq.~(\ref{Eq:cut_coulomb_pot}) 
can be tuned to achieve the best performance, but the default choice of $\gamma=5.0/R_{cut}$ Bohr$^{-1}$ and $R_{w}=1.092$ Bohr
works sufficiently well 
for all systems we have tested so far.

By replacing $1/|\bfr-\bfrp|$ by the truncated form $v^{cut}(|\bfr-\bfrp|)$ in Eq.~(\ref{Eq:bare_V_matr}), one obtains the truncated Coulomb matrix within the 
auxiliary basis, $V^{cut}_{\mu\nu}(\bfq)$, which is regular for $\bfq \rightarrow \bfzero$. 
A corresponding truncated screened Coulomb matrix 
$W^{cut}_0(\bfq,i\omega)$ can then be defined via Eq.~(\ref{Eq:screened_W_matr}) by replacing the full $V(\bfq)$ matrix 
by the truncated one $V^{cut}(\bfq)$ in the numerator. As described below, in doing so, one should be  
careful that the symmetrized dielectric function in the denominator of Eq.~(\ref{Eq:screened_W_matr}) should not be affected by the truncation procedure, i.e., it should
still be determined using the full Coulomb operator. For semiconductors and insulators, which are our concerns in the present work, 
the symmetrized dielectric 
function $\tilde{\varepsilon}$ matrix is finite and invertible everywhere in the BZ. The $\tilde{\varepsilon}$ matrix can be directly computed 
using Eq.~(\ref{Eq:symm_dielec_func}) for all $\bfq$ points except at $\bfq=0$. Because of the diverging behavior of $V_{\mu\nu}(\bfq)$ for $\bfq \rightarrow 0$,
as indicated by Eq.~(\ref{Eq:bare_V_symptotic}), the asymptotic behavior of the $\chi_0$ matrix (Eq.~(\ref{Eq:chi0_matrix})) needs to be taken care of 
in order to cancel the divergence in the $V$ matrix, similar to what is done in the plane-wave representation 
\cite{Baroni/Resta:1986,Hybertsen/Louie:1987}.

To this end, we adopt the scheme widely used in the context of the LAPW framework, which represents the dielectric function in terms of the eigenvectors
of the Coulomb matrix \cite{Friedrich/etal:2009,Friedrich/etal:2010,Jiang/etal:2013}. At $\bfq=0$, instead of diagonalizing the diverging full 
Coulomb matrix $V(\bfk=\bfzero)$, we diagonalize the truncated Coulomb matrix ${V}^{cut}(\bfk=\bfzero)$,
   \begin{equation}
     \sum_{\nu}{V}^{cut}_{\mu,\nu}(\bfk=\bfzero)X_{\nu,\lambda} = X_{\mu,\lambda} v_\lambda \, ,
   \end{equation}
where  $v_\lambda$ and $X_{\mu,\lambda}$ are its eigenvalues and its eigenvectors, which are real-valued for $\bfk=\bfzero$. 
Within its broad eigen-spectrum, there is one eigenstate standing out, with an eigenvalue that is significantly larger than all others, 
corresponding to the $G=0$ plane wave (i.e., the constant $1/\sqrt{\Omega}$ with $\Omega$ being the volume of the crystal). We shall denote 
this eigenstate as $\lambda=1$, and order the rest eigenstates according to their eigenvalues ($v_\lambda$ with $\lambda>1$) in an energetically descending manner. Naturally, $v_{\lambda=1}$ goes to infinity as $N_\bfk$ increases, whereby $V^{cut}$ 
approaches the full $V$.
Analyzing the components of the eigenvector $X_{\mu,1}$ reveals that this state has a predominant contribution from the nodeless $s$ functions  --  the only ABFs
representing non-zero net charges -- consistent with the nature of the $\bfG=0$ plane wave.  For the Coulomb matrix with a small $q$, we can diagonalize both the full
Coulomb matrix $V(\bfq)$ and its regular part $\bar{V}(\bfq)$ (cf. Eq.~(\ref{Eq:bare_V_symptotic})), and find that their eigenvectors are nearly the same. This is because the diverging part of the Coulomb matrix originates entirely from the $\lambda=1$ eigenvector, and removing this part from the Coulomb matrix amounts to shifting the value of $v_1$ downwards, but 
not changing the eigenvector $X_{\mu,1}$. Based on the above understanding, and inspired by the prescription of the LAPW basis set \cite{Draxl/Sofo:2006,Friedrich/etal:2009,Friedrich/etal:2010,Jiang/etal:2013}, we arrive at the following expression for the symmetrized dielectric function matrix within the Coulomb eigenvector basis representation,
\begin{widetext}
  \begin{align}
	  \tilde{\varepsilon}_{\lambda,\lambda'} (\bfq\rightarrow \bfzero, i\omega)=~~~~~~~~~~~~~~~~~~~~~~~~~~~~~~~~~~~~~~~~~~~~~~~~~~~~~~~~~~~~~~~~~~~~~~~~~~~~~~~~~~~~~~~~~~~~~~~~~~~~~~~~~~~~~~~~~~~~~~~~ & \nonumber \\
          \left\{ 
	     \begin{array}{ll} \displaystyle 1 - \frac{4\pi}{\Omega} \sum_{\bfk,\sigma}\left[ \sum_n 
                \frac{f'(\epsilon_{n,\sigma}^\bfk)|{\bf p}_{n,n,\sigma}^\bfk \cdot \hat{\bfq}|^2}{\omega^2} + \sum_{m <n}
		     \frac{2(f_{m,\sigma}^{\bfk}-f_{n,\sigma}^{\bfk})|{\bf p}_{m,n,\sigma}^\bfk \cdot \hat{\bfq}|^2}
		     {\left[(\epsilon_{m,\sigma}^{\bfk}-\epsilon_{n,\sigma}^{\bfk})^2 +\omega^2\right]
                         \left(\epsilon_{m,\sigma}^{\bfk}-\epsilon_{n,\sigma}^{\bfk} \right)} \right], & \lambda=\lambda'=1 \\
             \displaystyle -\sqrt{\frac{4\pi}{\Omega N_\bfk}}  \sum_{\bfk,\sigma} \sum_{m<n} \sum_{\nu}\frac{f_{m,\sigma}^{\bfk}-f_{n,\sigma}^{\bfk}}
                 {\epsilon_{m,\sigma}^{\bfk}-\epsilon_{n,\sigma}^{\bfk}}
                \left[ \frac{\left({\bf p}_{m,n,\sigma}^\bfk \cdot \hat{\bfq}\right)C_{n,m,\sigma}^\nu(\bfk,\bfk)}
             {\epsilon_{m,\sigma}^{\bfk}-\epsilon_{n,\sigma}^{\bfk} - i\omega} + c.c.\right]X_{\nu,\lambda'}\sqrt{v_{\lambda'}}, & \lambda=1, \lambda' > 1    \\
             \displaystyle -\sqrt{\frac{4\pi}{\Omega N_\bfk}}  \sum_{\bfk,\sigma} \sum_{m<n} \sum_{\mu} \sqrt{v_{\lambda}} X_{\mu,\lambda}
               \frac{f_{m,\sigma}^{\bfk}-f_{n,\sigma}^{\bfk}}{\epsilon_{m,\sigma}^{\bfk}-\epsilon_{n,\sigma}^{\bfk}}
               \left[\frac{C_{m,n,\sigma}^\mu(\bfk,\bfk) \left({\bf p}_{n,m,\sigma}^\bfk \cdot \hat{\bfq}\right)}
               {\epsilon_{m,\sigma}^{\bfk}-\epsilon_{n,\sigma}^{\bfk} - i\omega} + c.c. \right],     &  \lambda>1, \lambda' = 1    \\
             \displaystyle \delta_{\lambda,\lambda'} -\frac{1}{ N_\bfk}  \sum_{\bfk,\sigma} 
               \sum_{m<n} \sum_{\mu,\nu}\sqrt{v_{\lambda}} X_{\mu,\lambda}(f_{m,\sigma}^{\bfk}-f_{n,\sigma}^{\bfk})
               \left[ \frac{C_{m,n,\sigma}^\mu(\bfk,\bfk) C_{n,m,\sigma}^\nu(\bfk,\bfk)}
                           {\epsilon_{m,\sigma}^{\bfk}-\epsilon_{n,\sigma}^{\bfk} - \omega} + c.c.
               \right] X_{\nu,\lambda'}\sqrt{v_{\lambda'}} ,  & \lambda > 1, \lambda' > 1
             \end{array}\right.  \, 
        \label{Eq:dielectric_func_gamma}
  \end{align}
\end{widetext}
where ${\bf p}_{m,n,\sigma}^\bfk = \langle \psi_{m,\sigma}^\bfk| \hat{p} |\psi_{n,\sigma}^\bfk \rangle =
 -i\langle \psi_{m,\sigma}^\bfk| \nabla |\psi_{n,\sigma}^\bfk \rangle$ is the so-called momentum matrix, $\hat{\bfq} = \bfq/q$ is the unit vector along
the direction of $\bfq$, and $c.c.$ denotes complex conjugate. Furthermore, $f'(\epsilon)$ is the energy derivative of the Fermi-Dirac function,
          \begin{equation}
		  f'(\epsilon) = \delta f(\epsilon) / \delta \epsilon = -\frac{\mathrm{exp}\left((\epsilon-\mu)/\Delta)\right)}
		  {\Delta \left(1+\mathrm{exp}\left((\epsilon-\mu)/\Delta\right)\right)^2} 
          \end{equation}
which becomes a $\delta$-function, i.e., $-\delta(\epsilon-\mu)$, when the broadening parameter $\Delta$ 
(introduced to stabilize the calculations for metals or narrow-gap insulators) approaches zero. Equation~(\ref{Eq:dielectric_func_gamma}) indicates that
the head and wing terms [$\lambda=1$ or $\lambda'=1$ in Eq.~(\ref{Eq:dielectric_func_gamma})] of the dielectric function matrix 
in general depend on the direction along which $\bfq$ approaches zero. This is indeed the case
for anisotropic systems. The first term in the first line of Eq.~(\ref{Eq:dielectric_func_gamma}) corresponds to the so-called intraband contribution,
which is only present for metals. For insulators and semiconductors, this term is zero and doesn't need to be considered.

Now we have a computable formalism for the dielectric function in the basis representation of Coulomb eigenvectors. After its computation, we can transform
it back to the basis of ABFs,
           \begin{equation}
              \tilde{\varepsilon}_{\mu,\nu}(\bfq \rightarrow \bfzero, i\omega)  = \sum_{\lambda,\lambda'} X_{\mu,\lambda}  \, .
                            \tilde{\varepsilon}_{\lambda,\lambda'} (\bfq\rightarrow \bfzero, i\omega) 
                          X_{\nu,\lambda'}
           \end{equation}
The matrix form of the truncated screened Coulomb interaction $W^{cut}_{0,\mu,\nu}(q,i\omega)$ can then be computed in the entire BZ, 
and the $GW$ self-energy can be calculated by standard BZ sampling techniques.

Note that our scheme to deal with the Coulomb singularity in {\GnWn} calculations as outlined above 
is different from what is usually done in the literature. In the usual practice \cite{Massidda/Posternak/Baldereschi:1993,Jiang/etal:2013,Hueser/Olsen/Thygesen:2013,Wilhelm/Hutter:2017,Zhu/Chan:2020}, 
once the symmetrized dielectric function is properly treated
at $\bfq=0$, one can compute $\tilde{\varepsilon}^{-1}$ by performing a block-wise inversion and proceed to obtain 
the screened Coulomb interaction $W_0(\bfq, i\omega)$. $W_0(\bfq, i\omega)$ has a similar $1/q^2$ singularity behavior as
the bare Coulomb interaction $V(\bfq)$ for the so-called ``head" term, and additionally a $1/q$ singular behavior for the wing terms.
One can then subtract two analytic compensating functions from the integrand to smooth out these singular behaviors a
$\bfq \rightarrow 0$, and the BZ integration over these compensating functions can be done separately in an analytical way. 
In the end, the entire procedure boils down to performing an usual BZ summation (whereby the $\Gamma$-point can be omitted), 
and then adding two correction terms afterwards. Thus, the usual procedure can be viewed as the {\GnWn} analogy of 
the Gygi-Baldereschi scheme in HF calculations, whereas
our above-described scheme follows the spirit of the Alavi-Spencer scheme. A systematic comparison of the performance of
these two schemes is of high academic interest, but goes beyond the scope of the present work.


\section{\label{sec:convergence} Convergence tests}
  
Let us now investigate the precision and convergence behavior of our {\GnWn} implementation with respect to the 
numerical settings, including
the ABF basis set, the one-electron basis set, and the $\bfk$ point summation in 1BZ.

\subsection{\label{sec:conv_aux} Convergence test with respect to the number and shapes of ABFs}
As discussed in Sec.~\ref{sec:gw_lri}, our {\GnWn} implementation relies on the LRI approximation, whose accuracy further depends 
on the quality of the ABF basis set. In FHI-aims, the standard ABFs are constructed on the fly from the one-electron OBS employed in the preceding (g)KS calculations. As demonstrated in Ref.~\cite{Ihrig/etal:2015}, within LRI, 
typically a larger number of ABFs, especially those with higher angular momenta, is needed to achieve similar accuracy as 
in the standard RI-V scheme \cite{Ren/etal:2012}. As mentioned already in Sec.~\ref{sec:basis}, 
one practical way to improve the accuracy of LRI 
is to complement the OBS with additional functions of high angular momenta. The crucial point is that
these additional functions are only used to construct ABFs, but not used
in Eqs.~(\ref{eq:ao2mo}) or (\ref{Eq:KS_orbital}) to expand the eigen-orbitals. Namely, they don't enter the preceding
self-consistent KS-DFT calculations. 
In this way, the orbital products $\psi^\ast_{m,\sigma}(\bfr)\psi_{n,\sigma}(\bfr)$ on the left side of Eq.~(\ref{eq:RI_expan_mol}) 
does not change, but the $\left\{P_{\mu}\right\}$ set on the right side gets increased.  This can lead to improved accuracy of 
the expansion in Eq.~(\ref{eq:RI_expan_mol}), and thus the final results of the ground-state energy and/or quasiparticle band structure. In the FHI-aims input file, these
additional functions are labeled with a tag \textit{for\_aux}, signifying that these functions are only used to generate auxiliary functions. We follow the nomenclature of Ref.~\cite{Ihrig/etal:2015} by terming these additional \textit{for\_aux} functions 
as enhanced orbital basis set (OBS+).

In Ref.~\cite{Ihrig/etal:2015}, it has been shown that, for an OBS that contains at least up to $f$ functions, adding an additional 
5$g$ hydrogenic function to the OBS+ can yield an ABF set that is sufficiently accurate
for both HF and correlated methods like MP2 and RPA in molecular calculations. 
The shape and spatial extension of the 5$g$ hydrogenic function $u(r)$ depend on the effective
nuclear potential, $v(r)=Z/r$, with $Z$ being an effective charge, which governs the spatial extent of the solution of the 
radial Schr\"{o}dinger equation [Eq.~(\ref{Eq:radial_schrondinger})]. The smaller the value $Z$ is, the more extended 
the radial function $u(r)$.  It was found \cite{Ihrig/etal:2015} that the HF or MP2 results are not very sensitive to 
the precise value of $Z$, and an additional 5$g$ function defined by $Z=6.0$ is usually sufficiently good, yielding an accuracy that 
is comparable to the standard RI-V approximation.

Here we will examine the influence of the additional functions included in OBS+ on the band gaps of the periodic \GnWn calculations. 
To this end, we take Si and MgO crystals as 
the benchmark systems. The experimental lattice parameters ($a=5.431$~{\AA} for Si and $4.213$~\AA~ for MgO) are used in all calculations. Si is a covalently 
bonded semiconductor with a medium, indirect band gap, whereas MgO is an ionic crystal with a wide, 
direct band gap. 
Thus the two systems can be taken as representative examples for semiconductors and insulators, well suited for benchmark purposes. 

\begin{figure}[ht]
  \begin{picture}(200,340)(0,0)
     \put(100,245){\makebox(0,0){\includegraphics[width=0.4\textwidth,clip]{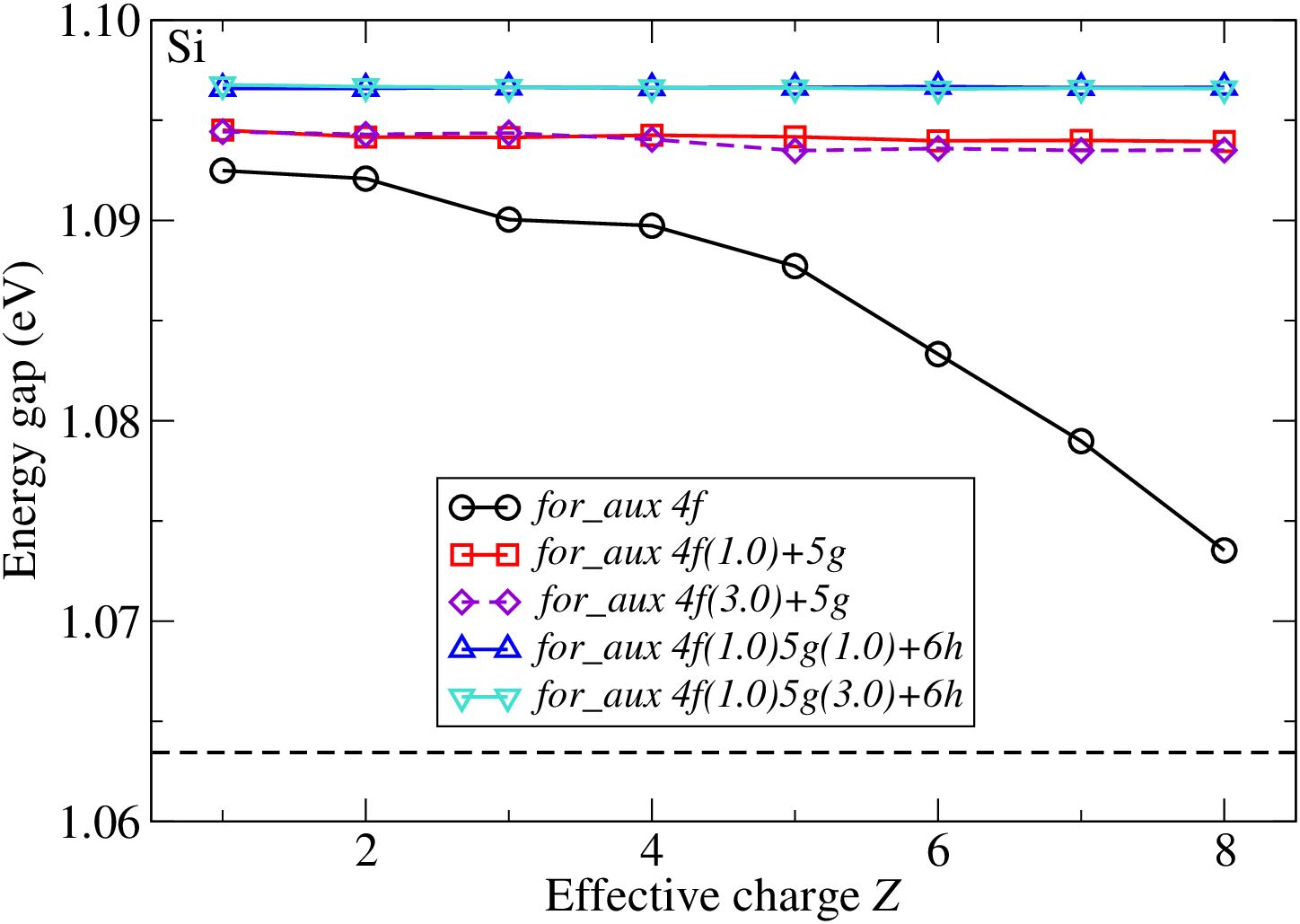}}}
     \put(100,80){\makebox(0,0){\includegraphics[width=0.4\textwidth,clip]{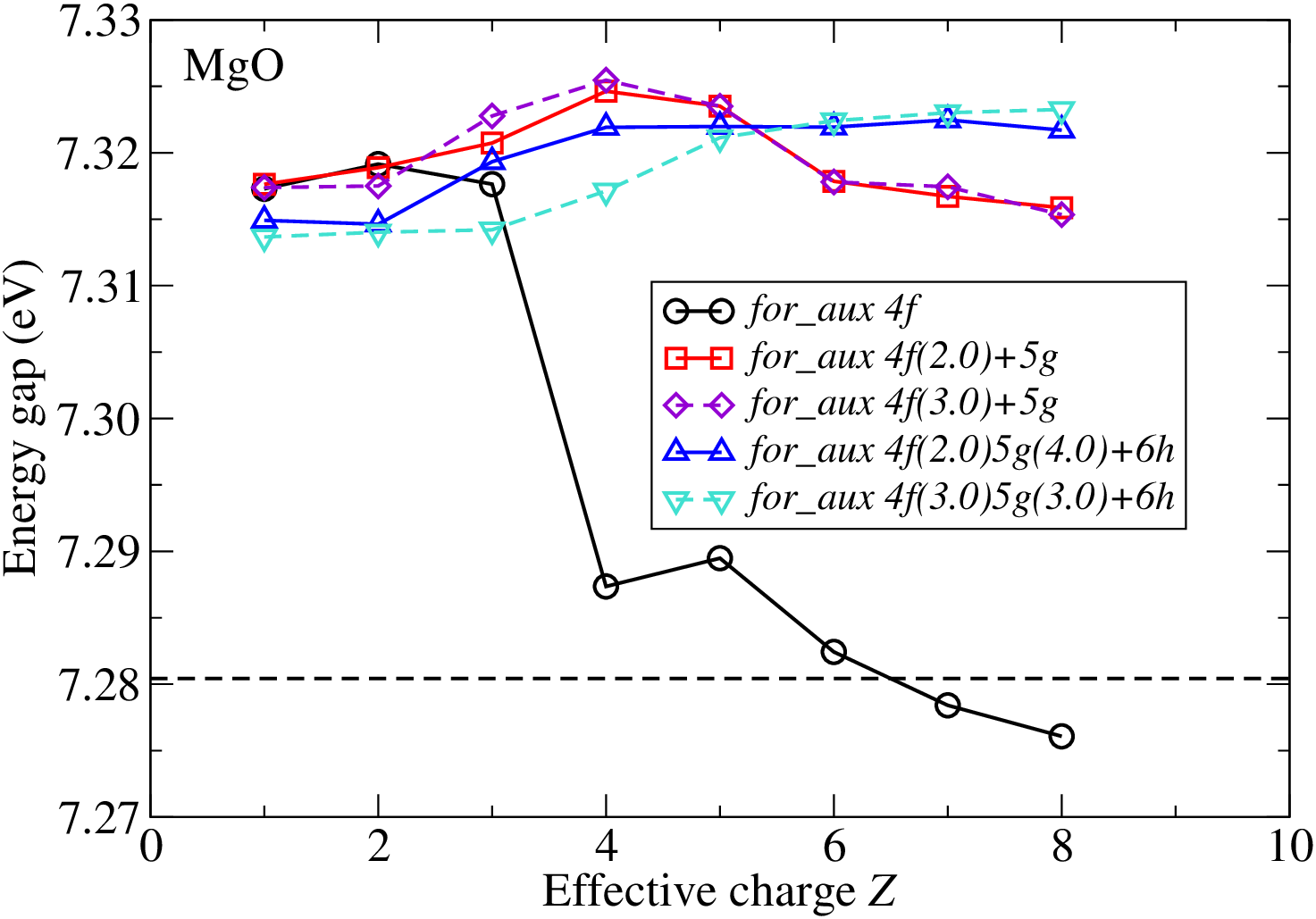}}}
     \put(20,330){\makebox(0,0){(a)}}
     \put(20,160){\makebox(0,0){(b)}}
  \end{picture}
\caption {The \GnWn@PBE band gaps for Si [panel (a)] and MgO [panel(b)] as a function of the effective charge $Z$ of the added hydrogen-like
	  \textit{for\_aux} function used to generate additional ABFs. The legend $4f(1.0)+5g$ means that in the OBS+, a $4f$ function with effective charge
	  $Z=1.0$ and a $5g$ function with varying effective charge $Z$ are added. Likewise, $4f(2.0)5g(4.0)+6h$ means a $4f$ with $Z=2.0$, a $5g$ function
	  with $Z=4.0$, and a $6h$ function with varying $Z$. The dash lines illustrate the behavior if the default $Z$ parameters
	  for $4f/5g$ \textit{for\_aux} functions are used.
          The FHI-aims \textit{tier} 2 basis and a $8\times 8 \times 8$ $\bfk$ grid were used in the preceding PBE calculations. The \GnWn band gaps 
	  are determined from full band structure calculations. The horizontal dash lines mark the results without including any  \textit{for\_aux} functions. }
\label{fig:for_aux_convergence}
\end{figure}

In Fig.~\ref{fig:for_aux_convergence} we present the calculated  {\GnWn} band gaps as a function of effective charge $Z$, for different 
 \textit{for\_aux} functions and their combinations included in OBS+. For the tests to be systematic, we have here checked not only 
 the influence of $5g$ functions but also of $4f$ and $6h$ functions.  The reference state for {\GnWn} calculations in 
Fig.~\ref{fig:for_aux_convergence} is generated from KS-DFT under the generalized gradient approximation (GGA) of Perdew, Burke, and Ernzerhof (PBE).
The FHI-aims \textit{tier} 2 basis and a $8\times 8 \times 8$ $\bfk$ grid were used in the PBE calculation. Note that the \textit{tier} 2 basis set contains basis functions up to $g$ for Si and O, and up to $f$ for Mg. 
The \GnWn band gap was determined from full band structure calculations along high symmetry lines in the BZ.

The \textit{for\_aux} functions are added in the following manner. We first add
a hydrogen-like $4f$ function, generated with an effective potential $Z/r$; 
by varying the value of $Z$, we check the influence of the shape of the
added $4f$ function on the calculated {\GnWn} band gap. Next we fix the $4f$ function with an ``optimal" $Z$ value, and add an additional $5g$
function governed by its own $Z$ value. 
Then we repeat this process by fixing both the $4f$ and $5g$ functions, and check the influence of an additional $6h$ function of varying spatial extent.
Our rule to choose the ``optimal" $Z$ value at each step is somewhat arbitrary, but here we pick the value in a window of $1.0 <=Z <=8.0$  
that gives the biggest increase of the band gap. 
Figure~\ref{fig:for_aux_convergence}(a) shows that, for Si, adding a $4f$ \textit{for\_aux} function can enlarge the calculated \GnWn gap roughly from 0.01 eV to
0.03 eV, depending on the chosen $Z$ value. Smaller $Z$ values (more extended $4f$ functions) tend to bring bigger corrections. Now, fixing the $4f$ function at
$Z=1.0$, and adding a $5g$ function in addition with varying $Z$ (denoted as $4f(1.0)+5g$), one can see that the band gap increases by about 2 meV, regardless the value of $Z$.  This is because once the $4f$ function is added, the largest part of the 
LRI error is removed, and the remaining error is not sensitive to the shape of the ABFs anymore. This is exactly the kind of effect 
we would like to achieve. Similarly, fixing $4f$ and $5g$ functions both at $Z=1$, and adding one $6h$ function 
(denoted as $4f(1.0)5g(1.0)+6h$), one gets a further 
increase of 2 meV of the band gap, independent of the $Z$ value of the $6h$ function. Such a convergence behavior with respect to 
the \textit{for\_aux} functions in 
OBS+ strongly suggests the remaining error arising from LRI is minor, and the \GnWn band gap for Si is well converged 
within 0.01 eV with respect to the ABFs. 

Similarly, for MgO, the change of the \GnWn band gap after adding a single $4f$ \textit{for\_aux} function (to both elements) is quite sensitive to the $Z$ value. 
As can be seen in Fig.~\ref{fig:for_aux_convergence}(b), the $4f$ functions
with smaller $Z$ values lead to bigger increases of the band gap (as large as 0.04 eV), while those with larger $Z$ values bring smaller increases
(or even slight decreases) of the band gap. Fixing the $4f$ function at $Z=2.0$ and adding a $5g$ function with varying $Z$, the band gap shows very little
change for small $Z$'s and a slightly larger increase for big $Z$ values. The overall effect is that the dependence of the obtained \GnWn band gap on the
shape of the $5g$ function is much reduced, although a remaining variation within a window of 0.01 eV can still be observed. Finally, fixing
$4f$ and $5g$ functions, and adding a $6h$ function with varying $Z$, the dependence of the obtained band gap on the $Z$ value of $6h$ function is further reduced. 
In contrast to the case of Si, for MgO the calculated band gap does not always get increased upon adding more \textit{for\_aux} functions. Instead, a slight 
decrease of the band gap can happen occasionally, indicating a more complex convergence behavior of the \GnWn calculation with respect to the ABFs. 
Nevertheless, the overall variation of the band gap values upon including additional $5g$ or $6h$ \textit{for\_aux} functions with varying $Z$ never exceed
a range of 0.02 eV. 
Such an uncertainty does not affect our subsequent convergence tests with respect to other computational parameters and the final benchmark calculations.

In the above tests, the chosen ``optimal" $Z$ values for $4f/5g$ \textit{for\_aux} functions are different for Si and MgO, but importantly, the
above discussed convergence behavior, as well as the final results are not sensitive to the actual $Z$ values over a wide range.
For example, using $4f(3.0)5g(3.0)$ instead of the ``optimal" $4f(1.0)5g(1.0)$ for Si and $4f(2.0)5g(4.0)$ for MgO, 
one only finds a difference of 0.2 meV for Si and 2 meV
for MgO in the calculated \GnWn@PBE band gap. The dash lines presented in Fig.~\ref{fig:for_aux_convergence} illustrate 
what happens if one starts with $4f(3.0)$ and $4f(3.0)5g(3.0)$  \textit{for\_aux} functions. 
Furthermore, adding $6h$ \textit{for\_aux} functions to the OBS+ leads to a factor of 2 increase of 
the computational cost but the accuracy gain is minor (the band gap changes below 0.01 eV). It should be noted that, in
our present implementation, the computational cost is only governed by the size of the auxiliary basis set, but not the shape
of the ABFs generated from the \textit{for\_aux} functions. That is, the computational cost is not affected by the $Z$ values
of the \textit{for\_aux} functions.
Under such circumstances, instead of using different settings for different systems,
we use a universal $4f(3.0)5g(3.0)$ \textit{for\_aux} basis setting in the following convergence tests and benchmark calculations 
for all materials.

\begin{figure}[t]
	\includegraphics[width=0.4\textwidth]{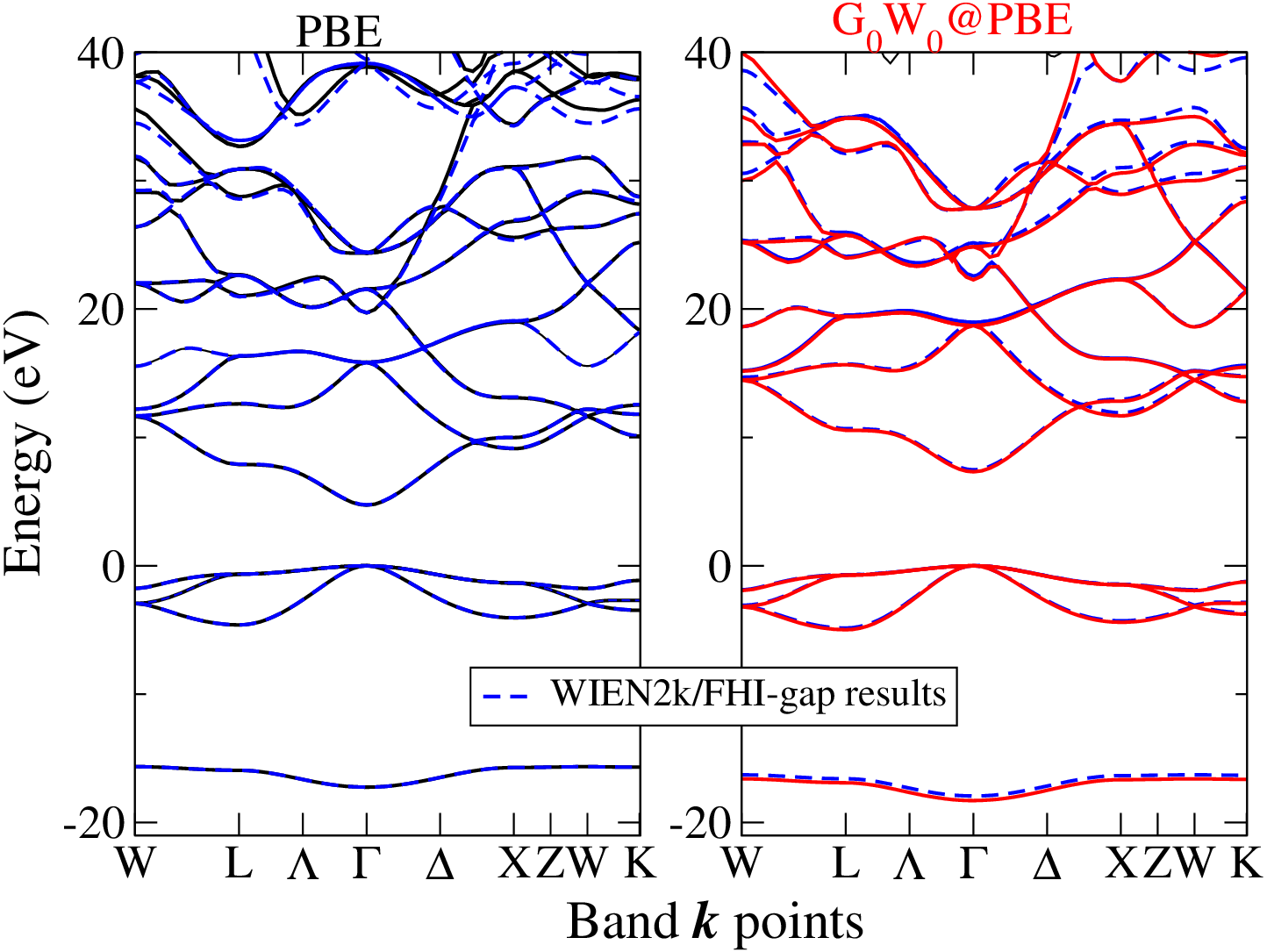}
	\caption{The PBE (left panel) and  \GnWn@PBE (right panel) band structures for MgO. Full lines are FHI-aims results (this work), obtained using \textit{tier} 2 basis set, the $4f(3.0)5g(3.0)$
	        \textit{for\_aux} functions, and a $8\times 8 \times 8$ $\bfk$ point mesh. The blue dash lines on the right panel
                       are the \GnWn band structure produced by the LAPW-based FHI-gap \cite{GomezAbal08,Jiang/etal:2013} code, based on
                       the PBE band structure determined by WIEN2k (blue dash lines on the left panel).}
			 \label{fig:MgO_FHI-aims-vs-FHI-gap}
\end{figure}

Finally, to demonstrate that the error incurred by LRI has indeed been made insignificantly small by adding \textit{for\_aux} functions as outlined above, 
we compare our calculated full \GnWn band structure to independent reference results
as obtained by the all-electron, LAPW-based FHI-gap code \cite{GomezAbal08,Jiang/etal:2013}, using MgO as an example.
In Fig.~\ref{fig:MgO_FHI-aims-vs-FHI-gap} the calculated band structure of MgO are presented for both PBE and {\GnWn}@PBE. 
The calculations employ the FHI-aims \textit{tier} 2 one-electron basis set, a fairly dense $8\times 8 \times 8$ $\bfk$ grid, and the above-noted $4f(3.0)5g(3.0)$ \textit{for\_aux} functions.
For comparison, the PBE and {\GnWn}@PBE band structures, as obtained respectively by 
the LAPW-based WIEN2k code \cite{Blaha/etal:2001} and the FHI-gap code, are shown as blue dash lines in
Fig.~\ref{fig:MgO_FHI-aims-vs-FHI-gap}. 
With its recent extension to complement the standard LAPW basis set with
high-energy local orbitals (HLOs) \cite{Laskowski/Blaha:2014} in its $GW$ calculations, the FHI-gap code, interfaced with
WIEN2k \cite{Blaha/etal:2001}, has been shown to deliver
highly accurate $GW$ band gaps \cite{Jiang/Blaha:2016} for a range of semiconductors and insulators. These include a set of 24 semiconductors and insulators that are typically used to benchmark the accuracy of theoretical approaches to electronic band structure of materials\cite{Jiang/Blaha:2016}, as well as copper and silver halides (CuX and AgX with X=Cl, Br, I) \cite{Zhang/Jiang:2019}, and several \textit{d}- and \textit{f}-electron oxides \cite{JiangH:2018}.
From Fig.~\ref{fig:MgO_FHI-aims-vs-FHI-gap}, it can be seen
that the FHI-aims results agree with those of FHI-gap very well over the entire BZ, not only for the valence-band maximum (VBM) and conduction-band minimum (CBM),
but also for quasiparticle energy levels much higher and lower in energy. Such close agreement between two 
very different {\GnWn} computational frameworks is remarkable. In fact, excellent agreement has also been achieved 
at the PBE level between the two all-electron codes -- FHI-aims and WIEN2k,
as can be seen in the left panel of Fig.~\ref{fig:MgO_FHI-aims-vs-FHI-gap}. More details can be found in 
Refs.~\cite{Lejaeghere/etal:2016,Huhn/Blum:2017}.  Furthermore, from Fig.~\ref{fig:MgO_FHI-aims-vs-FHI-gap}, we see
that with the standard NAO \textit{tier} 2 basis set, we can not only describe the occupied quasiparticle bands, 
but also the unoccupied bands up to 40 eV with acceptable accuracy.

As shown above, compared to Si, MgO is a somewhat more challenging system to control the error associated with LRI approximation,
used in our \GnWn implementation. Yet, excellent agreement between our NAO-based implementation (with an universal $4f(3.0)5g(3.0)$ \textit{for\_aux} function setting)
and the LAPW-based implementation -- the FHI-gap code can be achieved. 
A systematic benchmark study for a range of materials will be given in Sec.~\ref{sec:benchmark}. In the remaining part of this section, we continue to examine how our our \GnWn implementation converges with other numerical settings.

\subsection{\label{sec:conv_k} Convergence test with respect to {\bfk}-points}

In the above subsection, we discussed the convergence behavior of our \GnWn results with respect to the ABFs (more precisely the \textit{for\_aux} functions used to
generate additional ABFs), for a given $\bfk$ point mesh. With errors arising from the LRI approximation under control, 
we next examine how our \GnWn calculations converge with respect to the $\bfk$ point summation in evaluating the self energy. 

\begin{figure}[ht]
  \begin{picture}(200,340)(0,0)
     \put(100,245){\makebox(0,0){\includegraphics[width=0.4\textwidth,clip]{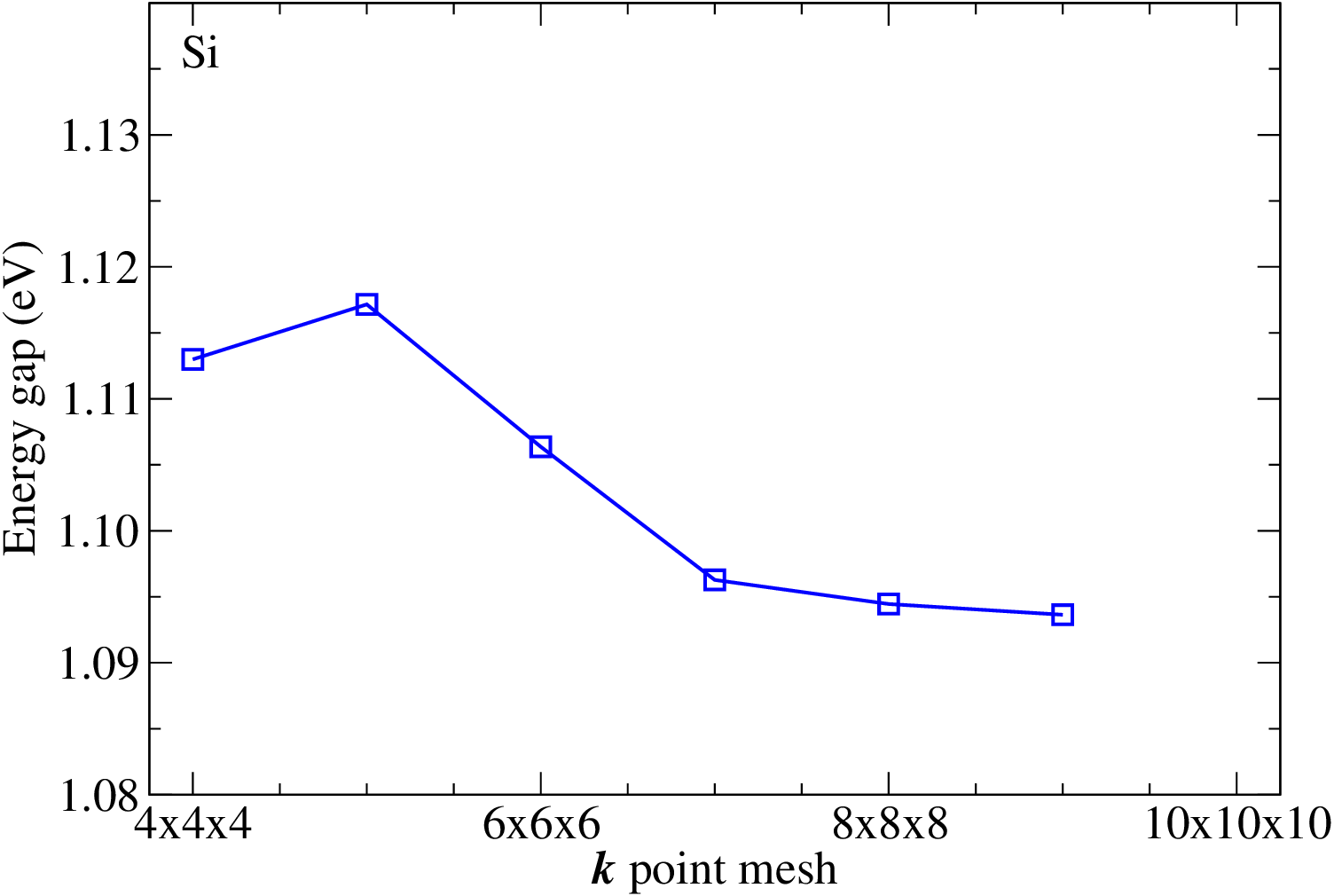}}}
     \put(100,80){\makebox(0,0){\includegraphics[width=0.4\textwidth,clip]{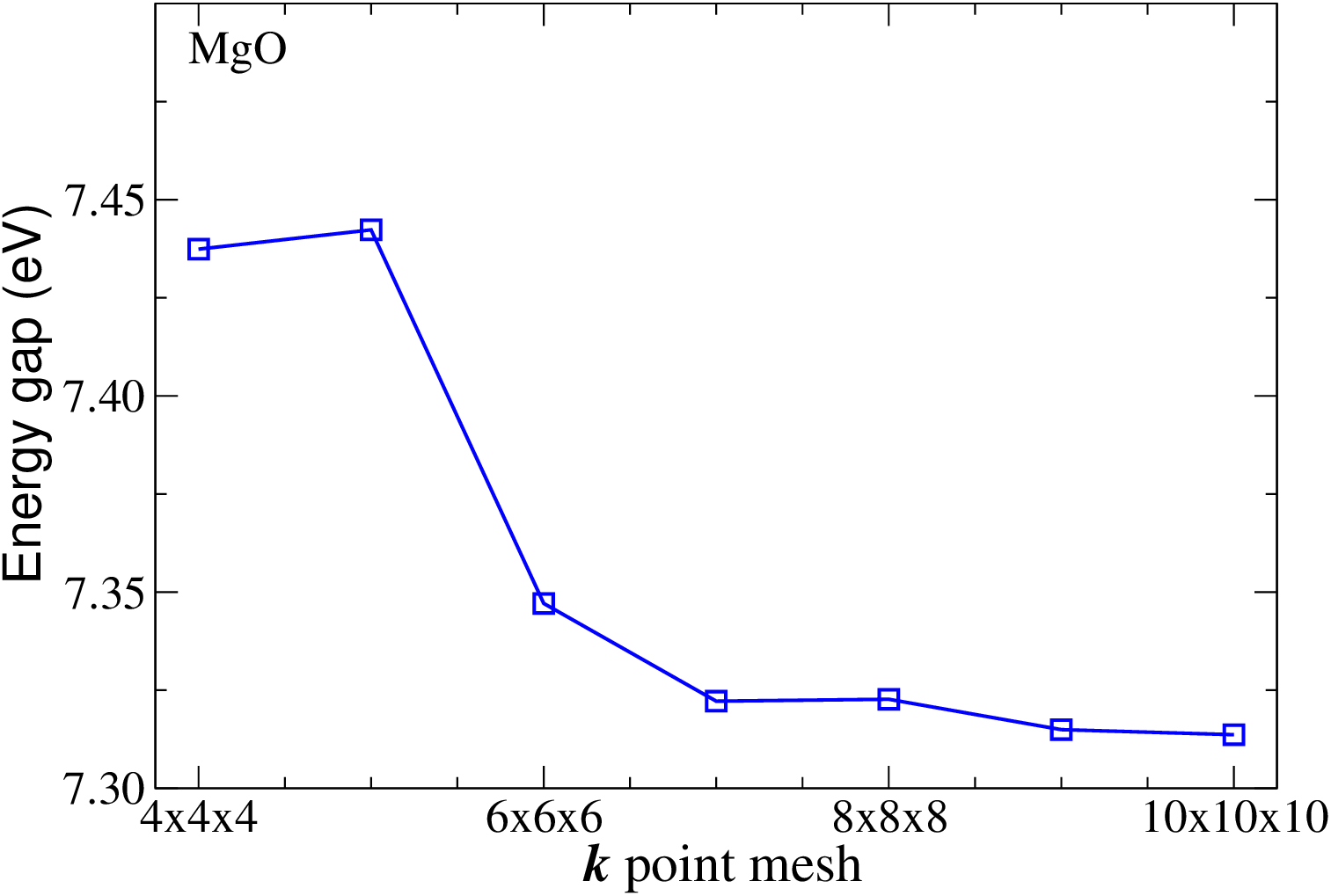}}}
     \put(20,330){\makebox(0,0){(a)}}
     \put(20,160){\makebox(0,0){(b)}}
  \end{picture}
\caption {The \GnWn@PBE band gaps for Si [panel (a)] and MgO [panel(b)] as a function of the $\bfk$ point mesh in the BZ summation.
	  A uniform mesh including the $\Gamma$ point is used. The FHI-aims \textit{tier} 2 is employed for the one-electron basis set, and the 
	  \textit{for\_aux} $4f(3.0)5g(3.0)$ functions are used to generate additional ABFs.}
\label{fig:kpoint_convergence}
\end{figure}

As discussed in Sec.~\ref{sec:gamma}, 
the summand in Eq.~(\ref{Eq:GW_selfe_element}) has an integrable divergence around $\bfq=0$, and the convergence behavior of a finite summation over $\bfq$
crucially depends on how we treat this singularity.  In Fig.~\ref{fig:kpoint_convergence}, we present the \GnWn@PBE band gap for Si and MgO as a function
of the $\bfk$ point mesh in the BZ summation. Calculations are done with \textit{tier} 2 one-electron basis set, and 
$4f(3.0)5g(3.0)$ \textit{for\_aux} functions. For Si, the calculated band gap varies smoothly as the
number of $\bfk$ points increases. Initially, it drops for increasing the $\bfk$ point density, but quickly saturates 
at around 1.09 eV for $\bfk$ meshes of $7\times 7 \times 7$ and denser. For MgO, the convergence of the band gap with respect to the $\bfk$ point mesh follows a similar behavior. The band gap essentially remains constant for a $\bfk$ point mesh beyond $7\times 7 \times 7$, and saturates at a value between 7.31 and 7.32 eV, with a tiny remaining variation of 0.01 eV. The origin of this additional complication is likely related to our special way of treating 
the $\Gamma$ point singularity: the truncated bare Coulomb operator together with an explicit evaluation of 
the dielectric function at $\bfq=0$ within the Coulomb eigenvector basis representation, as described
in Sec.~\ref{sec:gamma}. Investigating the origin
of this issue is beyond the scope of this paper, which focuses on the numerically converged $\bfk$ grid setting.

Below we will use $8\times 8 \times 8$ $\bfk$ point grid for further convergence studies with respect to other parameters and final benchmark calculations for crystals with a zinc-blende structure. For wurzite structure we use a $8\times 8\times 5$ $\bfk$ grid instead, due to the larger size of the unit cell along the $z$ direction.

\subsection{\label{sec:conv_basis} Convergence test with respect to one-electron orbital basis sets}

In the above two subsections, we have examined the convergence behavior of our \GnWn implementation with respect to 
the \textit{for\_aux} functions
and $\bfk$ point meshes for Si and MgO. We demonstrate that the numerical errors stemming from the LRI approximation and finite $\bfk$ point sampling can be well 
controlled and the uncertainties of the calculated \GnWn band gap due to these factors are within 0.01-0.02 eV. 
In this section we will check how our $\GnWn$ results converge with respect to the one-electron OBS. 
Here we emphasize again that the
one-electron OBS discussed in this subsection should be distinguished from the \textit{for\_aux} functions (included in OBS+) 
discussed in Sec.~\ref{sec:conv_aux}. While the former
is used to expand the SCF molecular eigen-orbitals that enter the subsequent \GnWn calculations, the latter (OBS+) is only employed 
to generate additional ABFs so as to reduce the numerical
error incurred by the LRI approximation, instead of improving the description of one-electron eigen-orbitals. 

The convergence to the complete basis set (CBS) limit using local atomic orbitals for correlated methods has long been considered a challenging problem. 
In quantum chemistry, experience has been gained to reach the CBS limit when using \textit{correlation consistent} or \textit{balanced} 
GTO basis sets \cite{Dunning:1989,Weigend/Ahlrichs:2005} combined with the Helgaker extrapolation procedure \cite{Helgaker/etal:1997}. 
Such a procedure works nicely for
small molecules, but cannot be directly applied to solids, due to a tendency to encounter linear dependence when 
using the standard GTOs in close packed structures. In this regard, NAOs 
are expected to be better behaved due to their strict locality in real space. Here
we will first examine how the \GnWn band gaps converge with respect to the standard
NAOs utilized in FHI-aims, and then further investigate the influence of highly localized Slater type orbitals (STOs).

\subsubsection{\GnWn calculations with one-electron NAO basis sets}
For NAO basis sets,
remarkable all-electron precision can be obtained for ground-state DFT calculations involving only occupied states \cite{Blum/etal:2009,Jensen/etal:2017}.
However, similar to the GTO case, the situation gets more involved for correlated calculations, like MP2 and RPA. The FHI-aims-2009 (``\textit{tier}") basis sets, 
originally developed for ground-state DFT calculations, can yield acceptably accurate results for MP2 and RPA binding energies if the
basis set superposition errors (BSSE) are corrected \cite{Ren/etal:2012}. 
Better accuracy can be achieved when using NAO-VCC-$n$Z basis sets \cite{IgorZhang/etal:2013}, via which results can be extrapolated to the CBS limit using
a two-point extrapolation procedure \cite{Helgaker/etal:1997}. 
However, it was found that the original NAO-VCC-$n$Z basis sets developed for molecules are not optimal for close-packed solids, due to their relatively larger radial extent (compared to the FHI-aims-2009 basis sets). Zhang \textit{et al.} \cite{IgorZhang/etal:2019} 
re-optimized these basis sets by removing the so-called ``enhanced minimal basis" and tightening the cutoff radius of the basis functions. The resultant basis sets, called localized NAO-VCC-$n$Z (loc-NAO-VCC-$n$Z) here, have been shown 
in Ref.~\cite{IgorZhang/etal:2019} to be able to yield accurate MP2 and RPA 
 binding energies for simple solids, with an appropriate extrapolation procedure.

\begin{figure}[ht]
  \begin{picture}(200,340)(0,0)
     \put(100,245){\makebox(0,0){\includegraphics[width=0.4\textwidth,clip]{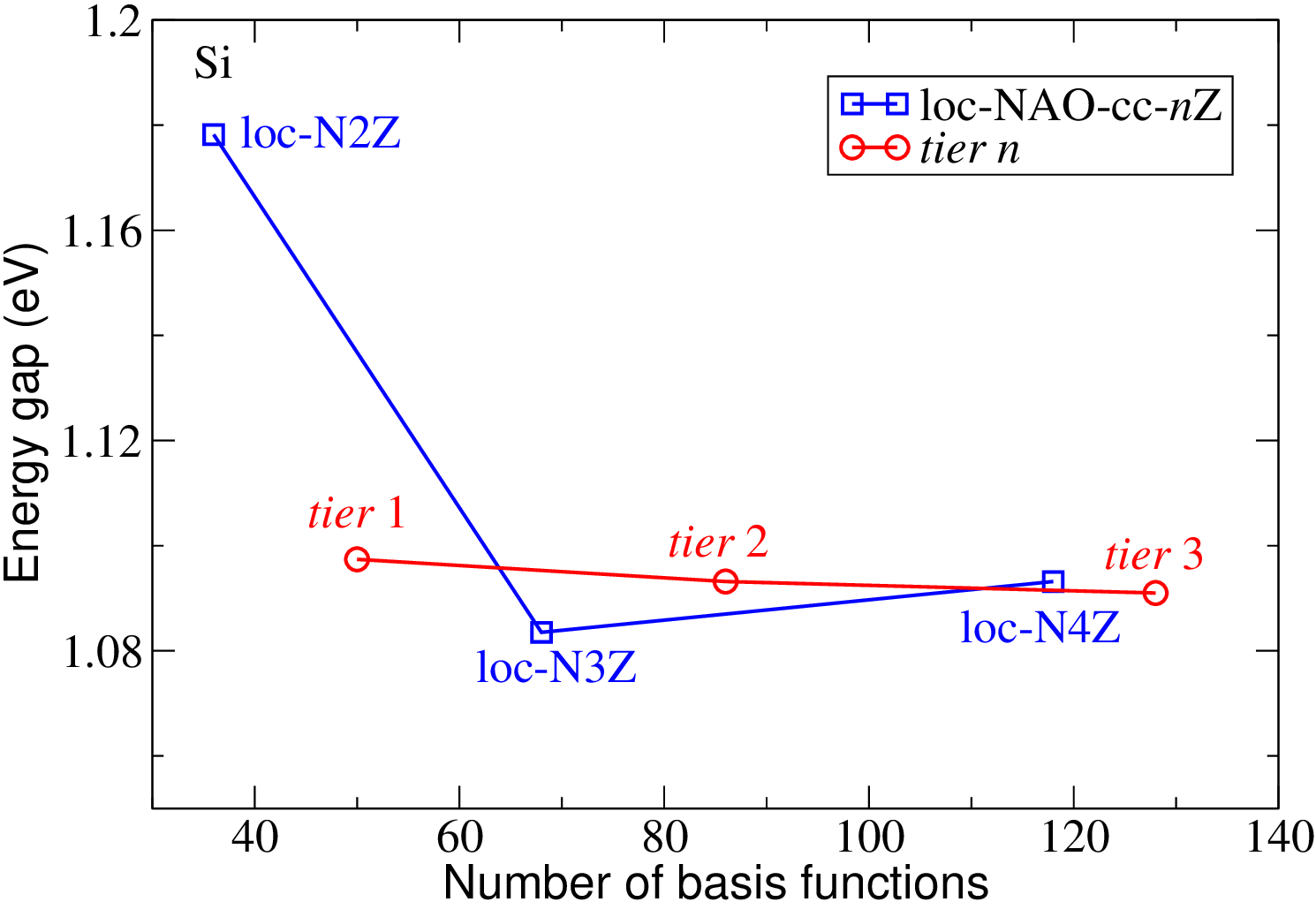}}}
     \put(100,80){\makebox(0,0){\includegraphics[width=0.4\textwidth,clip]{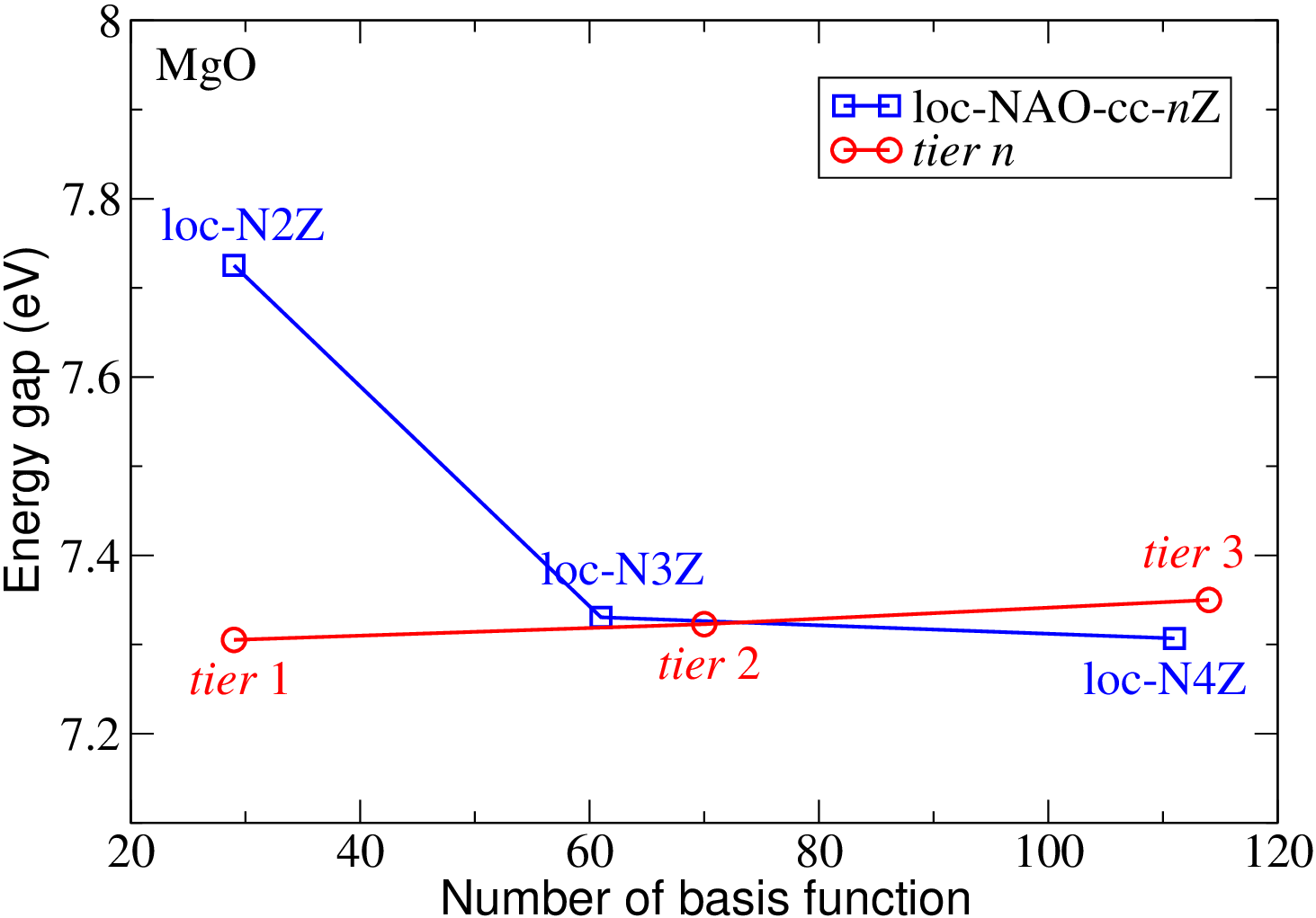}}}
     \put(20,330){\makebox(0,0){(a)}}
     \put(20,160){\makebox(0,0){(b)}}
  \end{picture}
\caption {The \GnWn@PBE band gaps for Si [panel (a)] and MgO [panel(b)] as a function of the size of the one-electron basis set 
	  per formula unit cell. Two types of hierarchical basis sets -- the loc-NAO-VCC-$n$Z (blue squares) and the FHI-aims-2009 (``\textit{tier}") 
	  basis sets (red circles) are used for the convergence test calculations. The ``loc-NAO-VCC-$n$Z" is abbreviated as ``loc-N$n$Z" to label
	  the data points.  A combined set of $4f(3.0)5g(3.0)$ hydrogen-like \textit{for\_aux} functions and 
	  a $8\times 8 \times 8$ $\bfk$ grid were used in the calculations. }
\label{fig:basis_convergence}
\end{figure}

The basis-set convergence of \GnWn energy levels is again different from MP2 and RPA for NAO basis sets. The FHI-aims-2009 
basis sets are usually used in FHI-aims \GnWn  
calculations for molecules. When using the \textit{tier} 4 basis set plus some additional diffuse GTOs, the \GnWn calculations
yield results comparable to those obtained with a very large aug-cc-pV5Z GTO basis sets \cite{Ren/etal:2012,Knight/etal:2016,Liuchi/etal:2020}. In the \GnWn case, the advantage of NAO-VCC-$n$Z basis sets is therefore less obvious, with a typically comparable or even slower convergence behavior.
Also, a reliable and easy-to-use procedure to extrapolate
\GnWn results to the CBS limit is not yet established. The different convergence behavior of NAOs for \GnWn and for
MP2/RPA is due to the fact that these methods are used to calculate different energetic properties. While \GnWn deals with 
individual states for which the error
due to inaccurate unoccupied orbitals is not significant, MP2 and RPA calculate total energies for which relatively small errors for individual states can
accumulate to an unmanageable level. 

Now we are at a stage to test the convergence behavior of both loc-NAO-VCC-$n$Z and FHI-aims-2009 basis sets for 
periodic \GnWn calculations. 
In Fig.~\ref{fig:basis_convergence} we present the calculated \GnWn band gaps for Si and MgO as a function of the basis set size. 
The $\bfk$ point mesh is fixed at $8\times 8 \times 8$ and the  $4f(3.0)5g(3.0)$ \textit{for\_aux} functions are used to 
generate additional ABFs. 
 For Si, the band gap only slightly decreases from 1.098 eV to 1.091 eV 
 from \textit{tier} 1 to \textit{tier} 3. On the other hand, when the loc-NAO-VCC-$n$Z basis sets are used, the band gap starts with a high value of 1.178 eV 
 with the loc-NAO-VCC-2Z basis, but quickly drops to 1.083 eV for the loc-NAO-VCC-3Z basis set and slightly increases back to 1.093 eV for 
 the loc-NAO-VCC-4Z basis set.  Thus, with the largest NAO-VCC-4Z and \textit{tier} 3 basis sets, we can achieve an satisfying agreement within 2 meV for the 
 \GnWn@PBE band gap. It is also interesting to see that the \textit{tier} 1 and \textit{tier} 2 basis sets can already yield band gaps that differ from the \textit{tier}
 3 result only by a few meV. From Fig.~\ref{fig:basis_convergence}(a) it can be seen that only the loc-NAO-VCC-2Z basis set appears to be out of the general trend.
 This is because, by eliminating the ``enhanced minimal basis", the loc-NAO-VCC-2Z is not yet sufficient to accurately describe the occupied states 
 in the preceding SCF calculations. Nevertheless, the overall convergence behavior indicates that, using either the FHI-aims-2009 or loc-NAO-VCC-$n$Z basis sets, 
 we can safely converge  the \GnWn gap for Si within 0.01 eV.
 
 For MgO, using the FHI-aims-2009 basis sets, the \GnWn@PBE gap increases from 7.31 eV to 7.35 eV. 
 When the loc-NAO-VCC-$n$Z basis sets are used, the
 calculated band gap quickly drops by about 0.4 eV from loc-NAO-VCC-2Z to loc-NAO-VCC-3Z. Further increasing the basis set to loc-NAO-VCC-4Z results in
 a slight decrease of the band gap from 7.33 to 7.31 eV. As such, the band gap values obtained with the \textit{tier} 2 and loc-NAO-VCC-$n$Z basis sets
 are already very close, but the difference becomes slightly larger when both types of basis sets get further increased. This behavior may reflect the presently achievable level of basis set convergence,
 since the basis set sizes of \textit{tier} 3 and loc-NAO-VCC-4Z are already rather large and the basis functions in a close-packed
 solids becomes linear dependent. In practical calculations, if the linear dependence becomes severe, we perform singular-value decomposition (SVD) to remove the redundant components, specifically 
 by eliminating the eigenfunctions of the overlap matrix below a threshold of $10^{-4}$. We note that the SVD procedure, while making 
 it possible to run calculations
 with large and linearly dependent basis sets, introduces some numerical uncertainty since the final \GnWn result will noticeably 
 depend on the choice of the 
 threshold value. This issue is more pronounced for MgO than Si, since MgO has a smaller lattice constant and Mg has a larger
 basis cutoff radius. 
 For example, when \textit{tier} 3 basis set is used, the \GnWn@PBE gap for MgO may range from 7.22 eV to 7.44 eV, depending on the choice of the SVD threshold. Such an uncertainty is too large to be acceptable. Thus for MgO, we take our \textit{tier} 2 result (7.32 eV) as the most accurate \GnWn@PBE gap value that our present numerical framework could offer.

\subsubsection{\label{sec:STOs}Complementing the one-electron NAOs with highly localized STOs}

With the standard NAOs, we face the problem that the quality of the convergence of the {\GnWn} results with respect to one-electron OBS cannot be rigorously assessed, because
systematically increasing the size of NAOs to the complete basis set limit is difficult due to their non-orthogonal nature. Already for MgO, the reliability of the results
obtained with \textit{tier} 3 or loc-NAO-VCC-4Z basis sets is limited by
the numerical instability arising from the linear dependence of the basis functions. In the LAPW framework, 
it has been observed that adding high-energy localized orbitals within the
muffin-tin sphere to the standard LAPW basis can significantly improve the band
gap values for materials such as ZnO \cite{Friedrich/Mueller/Bluegel:2011,Jiang/Blaha:2016,Nabok/Gulans/Draxl:2016}. 
Inspired by this experience, we test here
the possible impact by complementing the standard NAOs with highly localized orbitals.
As a first numerical test, highly localized Slater-type orbitals (STOs) are used. The reason to use STOs 
instead of NAOs here is that it is easier to control the spatial extent of
STOs by a single parameter. Furthermore, they can be deployed in an even-tempered way to systematically span the Hilbert space. 
The even-tempered STOs \cite{Raffenetti:1973} are given by
 \begin{equation}
     \phi_{\kappa,n,l,m}^\textnormal{STO} (\bfr) = {\cal N} r^{n-1}e^{-\zeta_{l,\kappa} r} Y_{lm}(\hat{\bfr})
 \end{equation}
 where ${\cal N}=(2\zeta)^{n}\sqrt{2\zeta/2n!)}$ is a normalization factor, and 
 \begin{equation}
     \zeta_{l,\kappa+1} = \alpha_l \beta_l^{\kappa-1} 2\zeta_{l,\kappa}, \kappa =1, 2, 3, \cdots \, .
 \end{equation}
That is, the exponents of the different STOs of the same $l$ follow a straight line on the logarithmic scale. 
In quantum chemistry, the even-tempered STOs and GTOs are the most popular 
choice for constructing systematic and efficient AO basis sets.
It can be shown that the overlap between two adjacent STOs or GTOs stays constant, leading to an even coverage of the Hilbert space \cite{Reeves/Harrison:1963,Cherkes/etal:2009}.
In this work, we employ 4 STOs for each $l$ channel, and for simplicity we set $n=l+1$
and $\beta_l=2$. Thus, for a given $l$, the STOs with $\zeta_{l,1}$ and $\zeta_{l,4}$ 
correspond to the most extended and the most localized functions, respectively. 
With these choices, the set of STOs is fully specified by the highest angular momentum $l_\textnormal{max}$ and the smallest component $\zeta_{l,1}=\alpha_l$ for
 $l<=l_\textnormal{max}$. The parameters $\alpha_{l}$ are chosen such that the overlaps between the STOs centering on 
 neighboring atoms are vanishingly small, and thus including them in the one-electrons OBS does not 
 cause numerical problems associated with linear dependencies.

\begin{figure}[ht]
\includegraphics[width=0.4\textwidth,clip]{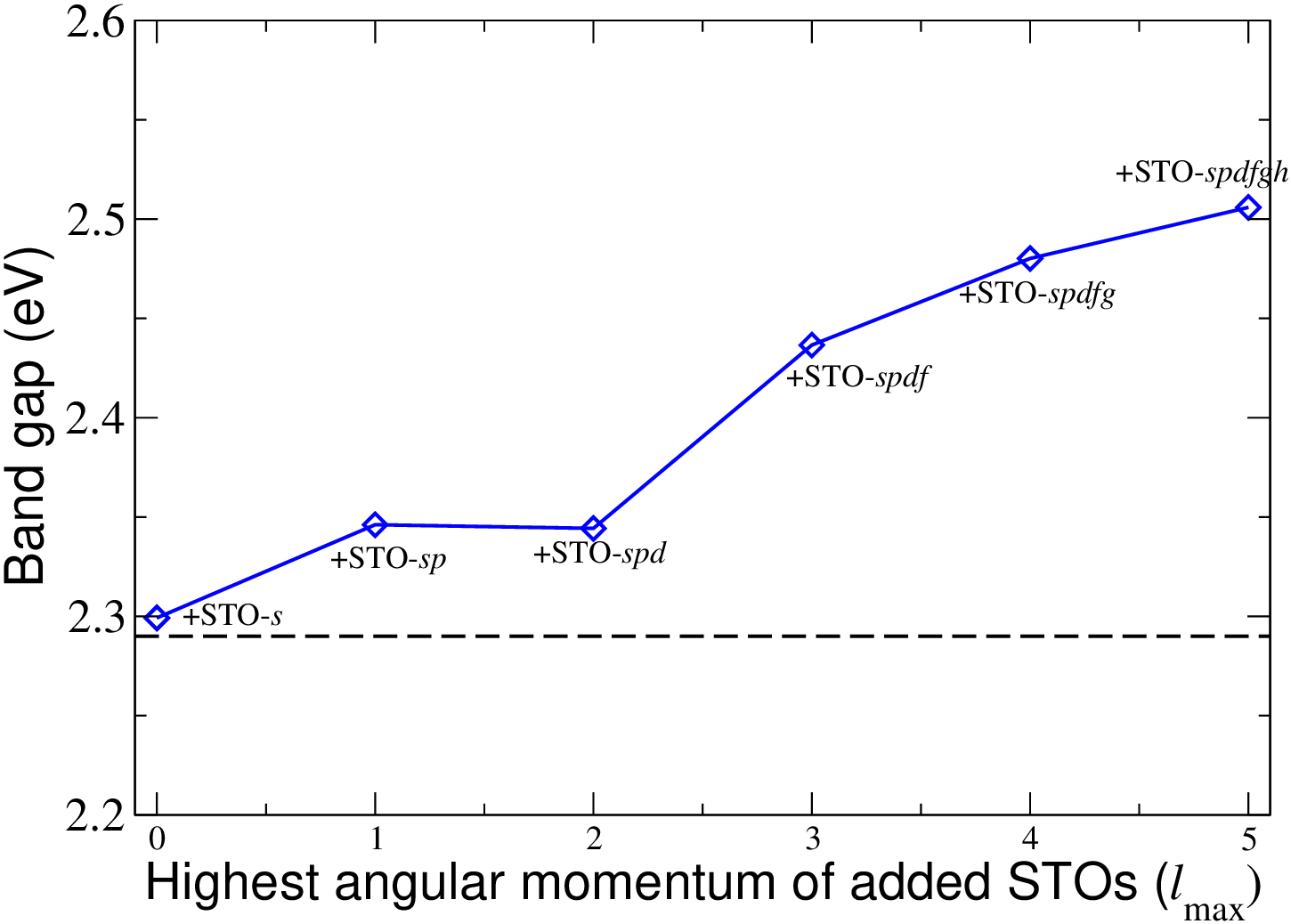}
\caption {
The variation of the \GnWn@PBE band gaps for zinc-blende ZnO upon adding STOs of increasing angular momenta to
FHI-aims-2009 NAO \textit{tier} 2 basis set. The chosen $\zeta_{l,1}$ values for $l=0,\cdots,5$ are $2.5,5,5,7.5,7.5,10$ for Zn and $5,10.10,10,15,20$ for O, respectively.
The horizontal dashed line marks the band gap value obtained with \textit{tier} 2 basis set.
A combined set of $4f(3.0)5g(3.0)$ hydrogen-like \textit{for\_aux} functions and 
a $8\times 8 \times 8$ $\bfk$ grid were used in the calculations. }
\label{fig:STO_convergence}
\end{figure}

In the following, we choose ZnO as the test example to check the possible influence of these highly localized STOs
in our NAO framework, since
previous experience accumulated in the LAPW community \cite{Friedrich/Mueller/Bluegel:2011,Jiang/Blaha:2016,Nabok/Gulans/Draxl:2016} showed that the localized orbitals have substantial impact on the band gap of ZnO. 
In Fig.~\ref{fig:STO_convergence}, we present the indirect \GnWn@PBE band gap for ZnO (in its zinc-blende structure)  obtained by adding STOs to the \textit{tier} 2 basis set. It can be seen that, as more STOs of increasing
angular momenta are included, the obtained band gap gradually increases, evolving from 2.29 eV obtained with the \textit{tier 2} basis set to 2.51 eV obtained with ``\textit{tier} 2 + STO-$spdfgh$" basis set (i.e., adding STOs with $l_\textnormal{max}=5$ to \textit{tier} 2). Although the band gap value is not
yet saturated for $l_\textnormal{max}=5$, the fully converged band gap is rather unlikely to exceed 2.6 eV, as can be
judged from the convergence plot shown in
Fig.~\ref{fig:STO_convergence}. Thus, this study suggests that complementing the standard NAO basis set with highly localized STOs enlarges the \GnWn band gap of ZnO by 0.2 to 0.3 eV. On the one hand, this
behavior is in qualitative agreement with what is found in the LAPW framework. On the other hand, the magnitude of
the correction brought out by the localized orbitals in the NAO framework is more than a factor of 2 smaller than the LAPW case.

We also performed a similar study as outlined above for Si and MgO. For these systems, adding the highly 
localized STOs up to $l_\textnormal{max}=5$ only gives rise to
a minor increase of the \GnWn@PBE band gap of just 0.04 eV. In Sec.~\ref{sec:benchmark}, we will
present benchmark band gap results for a set of crystals computed with both the standard \textit{tier} 2 
and the ``\textit{tier} 2 + STO-$spdfgh$"
basis sets. This will allow us to arrive at a more complete picture of the effects that localized orbitals can 
bring about in the NAO context. However, adding STOs up to $l_\textnormal{max}=5$ significantly
enlarges the size of one-electron basis set. For example, for ZnO the number of orbital basis functions 
($N_\textnormal{b}$) per unit cell drastically increases from 91 to 378, and this makes the entire calculation 20 times more expensive.

In a recent paper \cite{Zhu/Chan:2020}, Zhu and Chan showed that, with relatively small Dunning's cc-pVTZ or even cc-pVDZ GTO basis sets,
they can obtain fairly good band gaps for a range of insulating solids, including ZnO. Our preliminary tests with GTOs
indicate that the most extended (diffuse) functions in cc-pVDZ/TZ basis sets have to be excluded 
to circumvent the linear-dependent problem noted above.
On the other hand, when this is done, the obtained {\GnWn@PBE} band gap for ZnO with such modified 
cc-pVTZ basis set is 
on par with the more expensive ``\textit{tier} 2 + STO-$spdfgh$" basis set. This observation points 
to the possibility of developing compact and efficient NAO basis sets that are more suitable for 
all-electron {\GnWn} calculations. More investigations along this line are
needed to arrive at more faithful conclusions.

 \section{\label{sec:benchmark}Benchmark results}

In the previous section, we have examined the convergence behavior of the \GnWn band gaps with respect to
three numerical factors in our NAO-based implementation. Based on the above convergence tests, it appears that  $4f(3.0)5g(3.0)$ \textit{for\_aux} functions to generate extra ABFs guarantee a good accuracy under the LRI
approximation, and that a $8\times 8 \times 8$ $\Gamma$-inclusive $\bfk$ point grid is adequate for the BZ sampling
(for insulators) in cubic structures. Regarding the one-electron OBS, the FHI-aims-2009 \textit{tier} 2 basis set
seems to be adequate for ``simple" systems like Si or MgO, but for systems like ZnO, complementing \textit{tier} 2 with
highly localized STOs gives rise to to an increase of the band gap of 0.2 to 0.3 eV.
In this section, we will perform benchmark \GnWn calculations for a set of semiconductors and insulators. For these calculations,
the $4f(3.0)5g(3.0)$ \textit{for\_aux} functions are used throughout. As for the $\bfk$ grid, an $8\times 8 \times 8$ 
mesh is used for all crystals with cubic (zinc blende (ZB) for binary compounds) structure, whereas a reduced
$8\times 8 \times 5$ mesh is used for the wurzite (WZ) structure.
For one-electron OBS, both the ``\textit{tier} 2" and ``\textit{tier} 2 + STO-$spdfg$" basis set will be used. This
enables one to assess the overall influence of the highly localized orbitals on the computed \GnWn band gaps. Our 
results calculated in this work will be compared to the reference
values obtained using the FHI-gap code \cite{GomezAbal08,Jiang/etal:2013}.

\begin{table*}
  \centering
  \begin{tabular}{lccccccccc}
    \hline\hline
      \multirow{2}{*}{Crystals~~} & \multirow{2}{*}{Exp.($\Delta E_g^\text{ZPR}$)} & \multicolumn{3}{c}{~~FHI-aims~~} & ~~  & WIEN2K & ~~ & \multicolumn{2}{c}{FHI-gap} \\
                              \cline{3-5} \cline{7-7} \cline{9-10}
			    &                   &   ~~PBE~~  &  \GnWn(\textit{tier}2)   &\GnWn(\textit{tier}2+STO)  &  &  PBE & & \GnWn ($n_\text{LO}$=0) & \GnWn ($n_\text{LO}$=5)  \\
    \hline
	  AlAs     &  2.27(-0.039)   & 1.34(0.10)  &  2.03 & 2.23  & &  1.34(0.10) & & 1.94  & 2.06 \\ 
	  WZ-AlN & 6.44(-0.239) & 4.21  &  5.75 & 5.86 & &  4.14 & & 5.60 & 5.80 \\
	  AlP      &  2.53(-0.023)  & 1.58  &  2.32 & 2.43 & & 1.57 & & 2.25 & 2.37 \\
	  AlSb  &  1.73 (-0.039)  &  0.95(0.25) & 1.47 & 1.70 & & 1.03(0.22)  & & 1.40  & 1.50  \\
	  BAs      &  1.65(-0.151$^a$)    & 1.19  &  1.85  &1.91 & & 1.19 & & 1.78 & 1.83 \\
	  BN       &  6.66(-0.262$^b$)   & 4.46  &  6.32 & 6.31 & & 4.46 & & 6.04 & 6.36 \\
	  BP       &2.5,2.2 (-0.106$^d$) & 1.24  &  1.95 & 1.99 & & 1.34 & & 2.01 & 2.11 \\ 
	  C        & 5.85(-0.370)   & 4.13  &  5.61 & 5.60 & & 4.16 & & 5.49 & 5.69 \\ 
	  CdS      & ---     & 1.16  &  1.98 & 2.06 & & 1.14  & & 1.94 & 2.06  \\
	  WZ-CdS   & 2.64(-0.068)   & 1.17  &  1.97 & 2.05 & & 1.20  & & 2.02 & 2.19  \\
	  GaN      & 3.64(-0.173)  & 1.61  &  2.70 & 2.96 & & 1.68 & & 2.78 & 3.00 \\
	  GaP      & 2.43(-0.085) & 1.64  &  2.15 & 2.14 & & 1.66 & & 2.05 & 2.21 \\
	  LiCl     & 9.8 (-0.436$^d$)    & 6.33  &  8.55 &8.64 & & 6.30 & & 8.56 & 8.71  \\
	  LiF      & 14.48(-0.281$^b$)  & 9.20  & 13.79 & 13.49 & & 9.28 & & 13.36 & 14.27 \\
	  MgO      & 7.98(-0.154$^b$)  & 4.73  &  7.32 & 7.36 & & 4.75 & & 7.08 & 7.52 \\
	  NaCl     & 8.6(-0.098$^c$)    & 5.10  &  7.69 & 7.80 & & 5.12 & & 7.67 & 7.92 \\ 
	  Si      & 1.23 (-0.064)  & 0.61  &  1.09 & 1.13 & & 0.56 & &1.03 & 1.12 \\
	  SiC      & 2.57 (-0.145$^c$)  & 1.37  &  2.38 & 2.52  & & 1.36 & &2.23 & 2.38 \\
	  ZnO      & ---  & 0.69  &  2.29 & 2.51 & & 0.70 & & 2.05 & 2.78 \\
	  WZ-ZnO & 3.60(-0.156)   & 0.82  &  2.46  & 2.70 & & 0.83 & & 2.24 & 3.01 \\ \\
	  MD       &       &       &  -0.16 & -0.08 & &     & & -0.28 &  \\
	  MAD      &       &       &  ~0.17  & ~0.15 & &     & & ~0.28  & \\
    \hline\hline
    \multicolumn{5}{l}{
$^\text{a}$Theoretical estimates in Ref.~\cite{Bravic/Monserrat:2019}.} &
     \multicolumn{5}{l}{$^\text{b}$Theoretical estimates in Ref.~\cite{Antonius/etal:2015}.} \\
    \multicolumn{5}{l}{$^\text{c}$Theoretical estimates in Ref.~\cite{YimingZhang/etal:2020}.} & 
    \multicolumn{5}{l}{$^\text{d}$Theoretical estimates in Ref.~\cite{Shang/Yang:2020}.} \\
  \end{tabular}
 \caption{PBE and \GnWn@PBE band gaps calculated with FHI-aims, in comparison to the results obtained by FHI-gap \cite{GomezAbal08,Jiang/etal:2013} 
	  and the experimental values. The FHI-aims calculations were done with NAO \textit{tier} 2 and $4f(3.0)5g(3.0)$ \textit{for\_aux} basis functions. An
          $8\times 8\times 8$ $\bfk$ point mesh was used for all crystals except for wurzite structures, whereby a reduced $8 \times 8 \times 5$~ $\bfk$ grid was used.
          The WIEN2k PBE results and FHI-gap \GnWn@PBE results [with both standard LAPW basis ($n_\text{LO}$=0) and LAPW+HLOs basis ($n_\text{LO}$=5)]
           are mostly taken from Ref.~\cite{Jiang/Blaha:2016}. The mean deviation (MD) and mean absolute deviation (MAD)
           for the FHI-aims \GnWn results and the LAPW-based ($n_\text{LO}$=0) \GnWn results are obtained with reference
           to the LAPW+HLOs ($n_\text{LO}$=5)  results.
           The experimental values are directly cited from Ref.~\cite{Jiang/Blaha:2016}, originally taken from Refs.~\cite{Chiang/etal:1989,Madelung/etal:2004}.
           Systems with a prefix ``WZ" in their names mean the wurzite structure was used in the calculation; otherwise the 
           zinc blende structure was used instead.
           In all calculations the experimental lattice constants were used. The experimental band gaps
           are corrected for the ZPR effect, i.e., $E_g^\text{corr}=E_g(T=0)-\Delta E_g^\text{ZPR}$ with
           the ZPR contribution $\Delta E_g^\text{ZPR}$ given in parenthesis. Unless otherwise noted, 
           the ZPR term is estimated from a linear 
           extrapolation of the experimental gap $E_g(T)$ to $T=0$, as collected in Ref.~\cite{Cardona/Thewalt:2005}.   }
  \label{tab:gap_benchmark}
\end{table*}

\begin{figure}[t]
      \centering
      \includegraphics[width=0.48\textwidth]{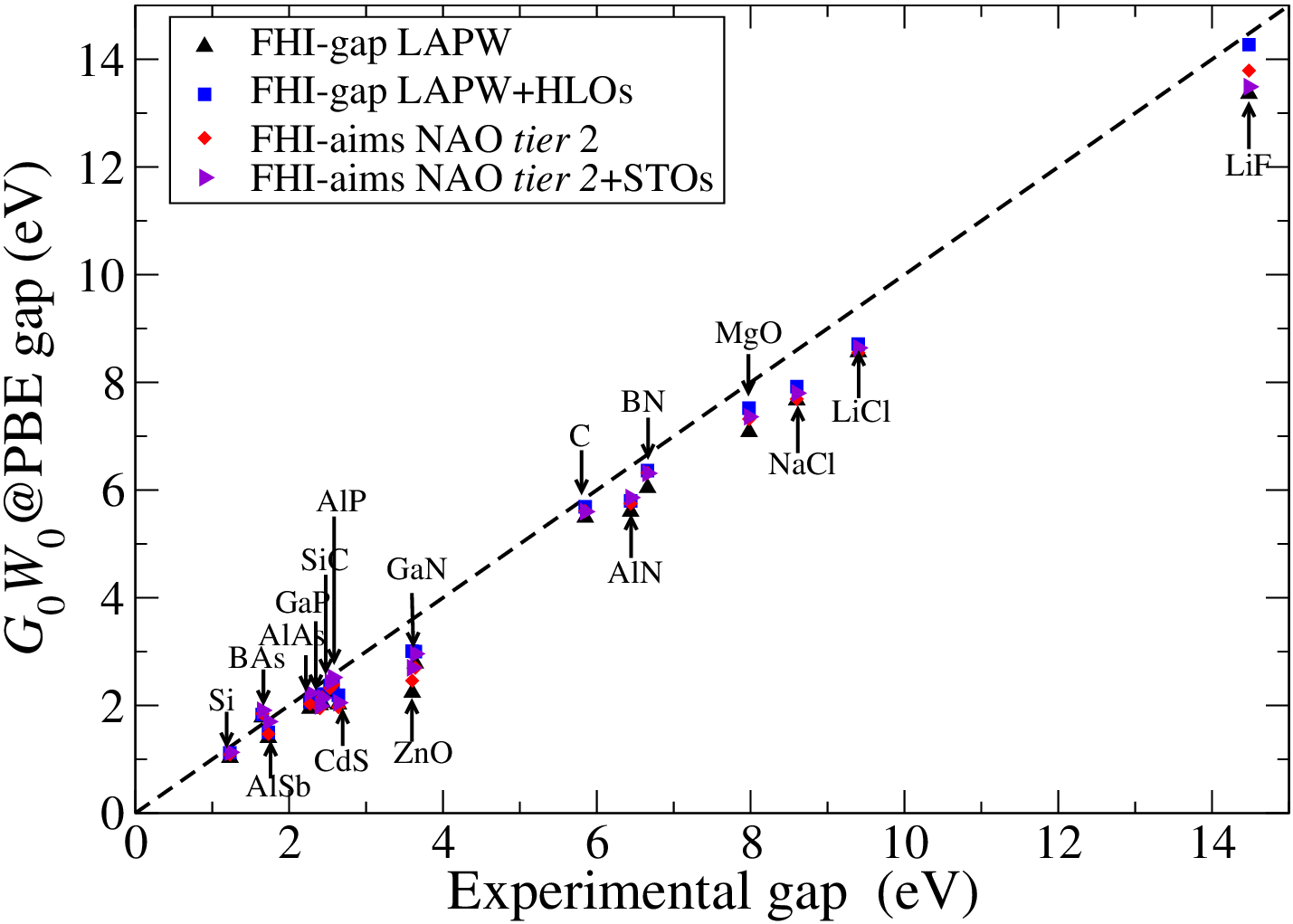}
     \caption{Calculated {\GnWn} band gaps versus the experimental ones for the materials presented in Table~\ref{tab:gap_benchmark}. 
	      For materials where results of both the ZB and WZ structures are calculated, only results for the WZ structure results are presented. The experimental values are corrected for the ZPR effect.} 
	    \label{fig:gw_gap}
\end{figure}

 In Table~\ref{tab:gap_benchmark} we present our calculated PBE and \GnWn@PBE band gaps of 21 crystals, covering systems from 
 small gap semiconductors 
 to wide gap insulators, formed with a wide variety of chemical elements. For AlN, CdS, and ZnO, results for both ZB and
 WZ crystal structures are shown. 
 Also presented are the reference PBE and \GnWn@PBE band gap results obtained respectively by the WIEN2k \cite{Blaha/etal:2001} and 
 FHI-gap \cite{GomezAbal08,Jiang/etal:2013} codes, as well as the experimental band gap values. The presented experimental values 
 are corrected for the zero-point renormalization (ZPR) effect, 
 \begin{equation}
   E_g^\text{corr}=E_g(T=0)-\Delta E_g^\text{ZPR}\, ,
 \end{equation}
 with $E_g(T=0)$
 being the experimental value extrapolated to zero temperature, and $\Delta E_g^\text{ZPR}$ being the ZPR contribution. 
 The calculated \GnWn@PBE gap values versus the experimental ones are further presented graphically in Fig.~\ref{fig:gw_gap}.
 The WIEN2k plus FHI-gap results are mostly taken from Ref.~\cite{Jiang/Blaha:2016}, except for CdS and LiCl, which are computed 
 for the first time in this work. The FHI-gap calculations were done with  $6\times 6 \times 6$ $\bfk$ grid, but tests show
 that the \GnWn band gaps are well converged with their own $\bfk$ point convergence strategy \cite{Jiang/etal:2013}. 
 For heavy elements where the spin-orbit coupling (SOC) effect is significant, both the PBE and \GnWn band gaps are 
 corrected by a SOC term given in parenthesis along with the PBE value. In FHI-aims calculations, the SOC effect 
 is treated using the non-self-consistent second variational method following a self-consistent scalar relativistic calculation \cite{Huhn/Blum:2017}.
 Again the PBE gaps show a remarkable level of agreement between FHI-aims and WIEN2k. 
 The FHI-gap \GnWn results were obtained both
 with the standard LAPW basis set ($n_\text{LO}$=0) and with the newly developed LAPW+HLOs prescription \cite{Laskowski/Blaha:2014,Friedrich/Mueller/Bluegel:2011}. In Ref.~\cite{Jiang/Blaha:2016}, 
 Jiang and Blaha demonstrated
 the influence of the addition of HLOs in the LAPW framework to the $GW$ band gaps of a sequence of materials. 
 It was shown that both the \GnWn and $GW_\text{0}$
 band gaps get enlarged when adding HLOs, but the magnitude of the correction varies from system to system, 
 ranging from less than 0.1 eV to more than 0.7 eV.

From Table~\ref{tab:gap_benchmark}, one may observe that, with few exceptions, the FHI-aims \GnWn@PBE results obtained with \textit{tier} 2 basis set
are found in between the FHI-gap results obtained respectively with  the standard LAPW basis set and with the highly converged LAPW+HLOs basis set. This behavior can also be clearly seen from the graphical presentation of the data in
Fig.~\ref{fig:gw_gap}. Actually, for about half of the materials, the FHI-aims \textit{tier} 2 results are fairly
close (within 0.1 eV) to the FHI-gap results with the LAPW+HLOs basis prescription, taken to be the reference here.
As indicated in Table~\ref{tab:gap_benchmark},  the mean deviation (MD) and mean absolute deviation (MAD) of the FHI-aims \textit{tier} 2
\GnWn results with respect to those obtained from the LAPW+HLOs($n_\text{LO}$=5) basis set (last column in Table~\ref{tab:gap_benchmark})  are substantially smaller than the corresponding MD and MAD values for the standard LAPW \GnWn results.
However, for systems where the addition of HLOs  
has a substantial influence, the FHI-aims \textit{tier} 2 results also show an appreciable underestimation of 
band gaps, as compared to the LAPW+HLOs results. This is particularly true for ZnO, LiF, NaCl, and GaN. A common feature of
these materials is that they have a strong ionic character and a direct band gap at the $\Gamma$ point. For ionic solids, 
the VBM and CBM originate from different chemical elements (and orbital characters) and hence may evolve differently as 
one increases the basis size, leading to an overall slower convergence of the band gaps. 

In particular, for ZnO, the \GnWn calculation is notoriously difficult to converge with respect to the unoccupied states. 
Since the remarkable finding of Shih 
\textit{et al.} \cite{Shih/etal:2010} that the earlier $GW$ calculations of ZnO were severely underconverged, different $GW$ implementations 
\cite{Stankovski/etal:2011,Friedrich/Mueller/Bluegel:2011,Klimes/Kaltak/Kresse:2014,Jiang/Blaha:2016,Nabok/Gulans/Draxl:2016,Rangel/etal:2020} have been tested for this system. 
Although differing in details, most $GW$ codes yield a substantial increase of the band gap for ZnO if the calculation is carefully converged with respect to
the unoccupied states. As already discussed above, within the LAPW framework, adding the HLOs in the LAPW framework increases the ZnO band gap by more than 0.7 eV 
\cite{Friedrich/Mueller/Bluegel:2011,Jiang/Blaha:2016,Nabok/Gulans/Draxl:2016}. Regarding the NAO basis set,  Table~\ref{tab:gap_benchmark} shows that, for both ZB- and WZ-ZnO, our \GnWn@PBE gap obtained with the FHI-aims-2009 \textit{tier} 2 basis set are larger by more than 0.2 eV than the corresponding values
obtained with the standard LAPW basis set, but are about 0.5 eV smaller than the values obtained with the fully converged 
LAPW+HLOs basis set.  
As demonstrated in Sec.~\ref{sec:STOs}, within our \GnWn implementation,
complementing \textit{tier} 2 with highly localized STOs leads to an enlargement of 0.2 to 0.3 eV of the band gap for ZB-ZnO.
In Table~\ref{tab:ZnO} we accumulate the \GnWn band gap values for ZnO in wurzite structure from various implementations, as
reported in the literature recently. One can see that, despite of the recent efforts, a scatter of 0.2 - 0.3 eV from
different \GnWn implementations is still visible.

\begin{table}
  \centering
  \begin{tabular}{lccccc}
    \hline\hline
     \multirow{2}{*}{Code} & \multirow{2}{*}{Framework} & & \multicolumn{2}{c}{\GnWn gap (eV)} & \multirow{2}{*}{Ref.} \\
                                   \cline{3-4} 
			    &                  & & @LDA   & @PBE  &    \\
    \hline
     FHI-aims  &  AE + NAO   & & 2.78  &  2.70  & this work  \\                    
     FHI-gap   &  LAPW       & &    ---   &  3.01 & \cite{Jiang/Blaha:2016}  \\
     J\"{u}liech   &  LAPW   & &  2.99  &  ---    & \cite{Friedrich/Mueller/Bluegel:2011}   \\  
     Exciting  &  LAPW       & & 2.94  &   ---    & \cite{Nabok/Gulans/Draxl:2016}     \\
     VASP      &  PAW + PW   & & 2.87  &  2.76     & \cite{Klimes/Kaltak/Kresse:2014}  \\
     Yambo     &  NCP + PW   & & 2.8   &  ---     & \cite{Rangel/etal:2020}   \\
     Abinit    &  NCP + PW   & & 2.8   &  ---     & \cite{Rangel/etal:2020}  \\
     BerkeleyGW &  NCP + PW  & & 2.8   &  ---     & \cite{Rangel/etal:2020}  \\
     PySCF      &  AE + STO  & & ---  &  3.08     & \cite{Zhu/Chan:2020}   \\
    \hline
    \end{tabular}
    \caption{\label{tab:ZnO}Recently reported \GnWn band gaps for WZ-ZnO from various implementations. In the ``Framework" column,
    ``AE" means "all-electron" and "NCP" means norm-conserving pseudopotential.} 
\end{table}

We then perform \GnWn@PBE calculations with the ``\textit{tier} 2 + STO-$spdfgh$" basis set for all systems and the obtained
results are also presented in Table~\ref{tab:gap_benchmark}. Close inspection reveals that the actual impact brought by the localized STOs varies from system to system. For most of the materials, adding the STOs to \textit{tier} 2 leads to an increase of the band gaps, similar to the LAPW+HLOs case. Among these, there are materials where the addition of STOs improves the FHI-aims \GnWn band gaps towards the ``reference" LAPW+HLOs values. This is in particular true for ZnO, NaCl, GaN, LiCl, and CdS where the improvement is substantial
($\sim 0.1$ eV or bigger). However, there are also cases (i.e., AlAs, AlP, AlSb, and SiC) where, after including the STOs,
the FHI-aims \GnWn band gap values exceed the LAPW+HLOs ones, and the agreement
between the two codes \textit{de facto}  deteriorates. Finally, we note that in a couple of cases, 
the addition of STOs leads to essentially no correction 
(C, BN, GaP) or a negative correction (LiF). Due to the varying impacts of the STOs, the MAD of the FHI-aims \GnWn band gaps
obtained with the ``\textit{tier} 2 + STO-$spdfgh$" basis set with reference to the LAPW+HLOs ones does not
show an appreciable improvement, although the MD value gets significantly reduced. 

In the $GW$ community, obtaining fully converged \GnWn results with respect to the one-electron OBS is of high academic
interest. In the NAO framework, complementing the standard NAOs with highly localized STOs seems to offer a
viable route towards this goal. However, a systematic addition of the STOs to the OBS, as described in
Sec.~\ref{sec:STOs}, leads to a very large number of one-electron basis functions, and as a consequence, the entire \GnWn calculation becomes one order of magnitude more expensive (9 times more expensive in case of NiO). 
This renders the combination of standard NAO with highly localized
STOs not a preferable scheme suitable for routine calculations.
Developing computationally more affordable schemes to correct the basis set incompleteness error for \GnWn calculations is
of current interest \cite{Betzinger/etal:2012,Betzinger/etal:2013,Betzinger/etal:2015,Loos/etal:2020}.
In particular, correction schemes that account for the response of the basis functions to the change of the effective potential can significantly improve the level of convergence towards the complete basis set limit
\cite{Betzinger/etal:2012,Betzinger/etal:2013,Betzinger/etal:2015}. This is because one effectively goes beyond 
the Hilbert space defined by the one-electron basis set by taking into account their variations upon responding to perturbations. 
For practical purposes, we suggest that the \textit{tier} 2 basis set be used as the default choice for periodic \GnWn 
calculations using FHI-aims. When numerically highly accurate results (band gap values converged within 0.1 eV or better) 
are needed, one can then check the quality of the obtained \GnWn results by further adding STOs to the OBS. 
Our preliminary investigations indicate that it is highly possible to 
develop more compact NAO basis sets for performing high-precision \GnWn calculations, especially
if the responses of basis functions can be efficiently included in the computational scheme.

From this benchmark study, one may also judge that, although the overall agreement of the two all-electron \GnWn implementations based on different numerical techniques is rather encouraging, the discrepancies for certain materials are
still too large to be acceptable. The level of agreement seen for the DFT-PBE band structures has not been achieved yet for \GnWn calculations. More investigations will be needed to clarify the possible origins.

Finally one may notice that $f$-electron materials are not included in the test set presented in Table~\ref{tab:gap_benchmark}. Our implementation is all-electron, and the core electron are included explicitly in the calculation. Hence the materials containing $f$-electrons do not pose formal difficulties. However, the convergence with respect to one-electron NAO basis set, as discussed above, will be even more challenging. We are currently testing our implementation on such materials, and the results will be reported in separate publications. The basis function response correction as mentioned above will be of great help there.
In this paper, we described in detail the formulation and algorithms of an all-electron periodic {\GnWn} implementation within the NAO framework. To our knowledge,
this is the first all-electron NAO-based {\GnWn} implementation that works with periodic boundary conditions. Our implementation was carried out within the 
FHI-aims code package \cite{Blum/etal:2009,Ren/etal:2012}.  With the achievement reported in this work, FHI-aims becomes a code that allows one to carry out both molecular 
and periodic \GnWn calculations, in an all-electron fashion, within a unified numerical framework. We performed systematic convergence tests, and identified a set of computational
parameters that can be used as default settings in \GnWn calculations to obtain reliable results. With such a default setting, we benchmarked our implementation
by computing \GnWn@PBE band gaps for a set of semiconductors and insulators, and compared the obtained band gaps to the independent
Wien2k plus FHI-gap \cite{GomezAbal08,Jiang/etal:2013} 
results. We found that, with the standard NAO \textit{tier} 2 basis, one can obtain band gaps that are already in fairly 
good agreement with those obtained by FHI-gap with highly 
accurate LAPW+HLOs basis set. Challenging situations do exist, like the famous ZnO example, 
where NAOs suffer from a similar one-electron basis under-convergence issue as other basis frameworks do, though to a somewhat less extent. Complementing the FHI-aims-2009 \textit{tier} 2 NAO basis set with highly localized STOs, one obtains 
appreciable improvements of the band gap value for a fraction of the materials, including ZnO.  However, the computational
scheme by complementing NAOs with highly localized STO is extremely expensive, and is not suitable for routine calculations.
For most practical purposes, we recommend to use the FHI-aims-2009 NAO \textit{tier} 2 basis set, which delivers useful
accuracy. Importantly, even this level of \GnWn calculations can significantly improve over the computationally cheaper but more empirical hybrid density functional approximations. While hybrid DFT calculations are much more readily convergable to the complete basis set limit, they cannot be parameterized to fully represent the local electronic structure variations in hybrid materials with conceptually different components (e.g., organic-inorganic interfaces). In contrast, $\GnWn$, with practically affordable basis set, can provide a much more uniform level of theory that encompasses the locally relevant screening effects in a natural way.

The compactness and locality of NAOs in principles enable efficient periodic GW implementations, allowing for tackling complex materials at reduced cost. Our benchmark calculations show that high-quality GW results can be obtained using modest NAO basis sets, except for some challenging systems. Generating compact NAO basis sets that can efficiently represent unoccupied Hilbert space is highly desirable, but requires considerably more efforts that goes beyond the scope of the present work. In this endeavor, independent implementations using systematic basis sets such as LAPW+HLOs will provides invaluable reference results for benchmark purposes. 

Our implementation is massively parallel. Due to the local nature of our basis set, our
implementation can be readily applied to 1D and 2D systems. Benchmark calculations for the efficiency and scalability of our
implementation, as well as its performance for systems with lower dimensions, will be presented in a future paper. Finally we are also working to extend our implementation to treat metallic systems.

\section{\label{sec:conclusion} Conclusion and outlook}
In this paper, we described in detail the formulation and algorithms of an all-electron periodic {\GnWn} implementation within the NAO framework. To our knowledge,
this is the first all-electron NAO-based {\GnWn} implementation that works with periodic boundary conditions. Our implementation was carried out within the 
FHI-aims code package \cite{Blum/etal:2009,Ren/etal:2012}.  With the achievement reported in this work, FHI-aims becomes a code that allows one to carry out both molecular 
and periodic \GnWn calculations, in an all-electron fashion, within a unified numerical framework. We performed systematic convergence tests, and identified a set of computational
parameters that can be used as default settings in \GnWn calculations to obtain reliable results. With such a default setting, we benchmarked our implementation
by computing \GnWn@PBE band gaps for a set of semiconductors and insulators, and compared the obtained band gaps to the independent
Wien2k plus FHI-gap \cite{GomezAbal08,Jiang/etal:2013} 
results. We found that, with the standard NAO \textit{tier} 2 basis, one can obtain band gaps that are already in fairly 
good agreement with those obtained by FHI-gap with highly 
accurate LAPW+HLOs basis set. Challenging situations do exist, like the famous ZnO example, 
where NAOs suffer from a similar one-electron basis under-convergence issue as other basis frameworks do, though to a somewhat less extent. Complementing the FHI-aims-2009 \textit{tier} 2 NAO basis set with highly localized STOs, one obtains 
appreciable improvements of the band gap value for a fraction of the materials, including ZnO.  However, the computational
scheme by complementing NAOs with highly localized STO is extremely expensive, and is not suitable for routine calculations.
For most practical purposes, we recommend to use the FHI-aims-2009 NAO \textit{tier} 2 basis set, which delivers useful
accuracy. Importantly, even this level of \GnWn calculations can significantly improve over the computationally cheaper but more empirical hybrid density functional approximations. While hybrid DFT calculations are much more readily convergable to the complete basis set limit, they cannot be parameterized to fully represent the local electronic structure variations in hybrid materials with conceptually different components (e.g., organic-inorganic interfaces). In contrast, $\GnWn$, with practically affordable basis set, can provide a much more uniform level of theory that encompasses the locally relevant screening effects in a natural way.

The compactness and locality of NAOs in principles enable efficient periodic GW implementations, allowing for tackling complex materials at reduced cost. Our benchmark calculations show that high-quality GW results can be obtained using modest NAO basis sets, except for some challenging systems. Generating compact NAO basis sets that can efficiently represent unoccupied Hilbert space is highly desirable, but requires considerably more efforts that goes beyond the scope of the present work. In this endeavor, independent implementations using systematic basis sets such as LAPW+HLOs will provides invaluable reference results for benchmark purposes. 

Our implementation is massively parallel. Due to the local nature of our basis set, our
implementation can be readily applied to 1D and 2D systems. Benchmark calculations for the efficiency and scalability of our
implementation, as well as its performance for systems with lower dimensions, will be presented in a future paper. Finally we are also working to extend our implementation to treat metallic systems.

\section*{Acknowledgments}
We would like to thank Igor Ying Zhang, Patrick Rinke, and Christian Carbogno for helpful discussions. XR acknowledges the funding supports from Chinese National Science Foundation (Grant Nos. 11574283 and 11874335) and the Max Planck Partner Group Project. This work also received funding from the European Union's Horizon 2020 Research and Innovation Program, The NOMAD CoE (No. 951786), and ERC Advanced Grant, TEC1P (No. 740233).


\section*{References}
\bibliography{./CommonBib}

\end{document}